\documentclass[journal,comsoc]{IEEEtran}

\usepackage[T1]{fontenc}
\usepackage{amsmath}
\usepackage{graphicx}
\usepackage[cmintegrals]{newtxmath}
\usepackage{subcaption}
\usepackage{multirow}
\usepackage{booktabs}
\usepackage[ruled]{algorithm2e}
\usepackage{enumitem}
\newcommand{\RN}[1]{%
  \textup{\uppercase\expandafter{\romannumeral#1}}%
}

\begin{document}

\title{Novel Architectures for Unsupervised Information Bottleneck based Speaker Diarization of Meetings}

\author{Nauman~Dawalatabad,~\IEEEmembership{Student Member,~IEEE,}
        Srikanth~Madikeri,~\IEEEmembership{Member,~IEEE,} \\
       C. Chandra~Sekhar,~\IEEEmembership{Member,~IEEE,}
        and~Hema~A.~Murthy,~\IEEEmembership{Senior Member,~IEEE,}   
\thanks{N. Dawalatabad, C. C. Sekhar and H. A. Murthy are with the Indian Institute of Technology Madras, India. (e-mail: \{nauman, chandra, hema\}@cse.iitm.ac.in).} 
\thanks{S. Madikeri is with the Idiap Research Institute, Martigny, Switzerland. (e-mail: srikanth.madikeri@idiap.ch).}}  

\maketitle

\IEEEpubid{\begin{minipage}{\textwidth}\  \vspace{1cm}  \\[12pt] \centering 
  This article has been accepted for publication in IEEE/ACM TRANSACTIONS ON AUDIO, SPEECH, AND LANGUAGE PROCESSING.~\copyright2020 IEEE.
\end{minipage}}

\begin{abstract}

 Speaker diarization is an important problem that is topical, and is especially useful as a preprocessor for conversational speech related applications.  
The objective of this paper is two-fold: (i)  segment initialization by uniformly distributing speaker information across the initial segments, 
and (ii) incorporating speaker discriminative features within the unsupervised diarization framework.
In the first part of the work, a varying length segment initialization technique for Information Bottleneck (IB) based speaker diarization system using phoneme rate as the side information is proposed.
This initialization distributes speaker information uniformly across the segments and provides a better starting point for IB based clustering. 
In the second part of the work, we present a Two-Pass Information Bottleneck (TPIB) based speaker diarization system that incorporates speaker discriminative features during the process of diarization.
The TPIB based speaker diarization system has shown improvement over the baseline IB based system.    
During the first pass of the TPIB system,  a coarse segmentation is performed using IB based clustering.
The alignments obtained are used to generate speaker discriminative features using a shallow feed-forward neural network and linear discriminant analysis.  
The discriminative features obtained are used in the second pass to obtain the final speaker boundaries. 
In the final part of the paper, variable segment initialization is combined with the TPIB  framework. 
This leverages the advantages of better segment initialization and speaker discriminative features that results in an additional improvement in performance.
An evaluation on standard meeting datasets shows that a significant absolute improvement of 3.9\% and 4.7\%  is obtained on the NIST and AMI datasets, respectively.

\end{abstract}


\IEEEpeerreviewmaketitle

\section{Introduction}
\label{sec:intro}

\IEEEPARstart{G}{iven} an audio signal, \textit{speaker diarization} involves answering the question of \textit{``Who spoke When?"} \cite{xavier12-review}. A speaker diarization system annotates audio with relative speaker labels. The task involves estimating the number of speakers and assigning speech segments to different speakers. Speaker diarization has been used in various domains, such as telephone conversations, broadcast news, and meetings \cite{xavier12-review}. Diarization of conversational audio meetings is considered to be a challenging task owing to the spontaneity in the conversation. Diarization systems are often used as front-ends in applications that include automatic speech recognition, spoken keyword spotting, and speaker recognition \cite{ibmwatson}.

Information Bottleneck (IB) based approach to speaker diarization has shown competitive performance for meeting recordings \cite{deepu09-ib,deepu09-klRealign,deepu11-fuseTDOA}. Owing to its non-parametric nature, IB based diarization has a very low Real Time Factor (RTF) value \cite{deepu09-ib,srikanth15-improveRuntime}. RTF is the time taken by a system to process 1 second of speech data.
Since diarization systems are mostly used in the pre-processing stage of many conversational speech applications, it is desirable to have diarization systems with low RTF value.

In general, an approach to unsupervised diarization of a conversational speech includes speech segment initialization followed by the bottom-up agglomerative clustering of the segments \cite{xavier12-review}. 
The major challenges in building an unsupervised speaker diarization system lie in the initialization of segments for clustering, to obtain speaker discriminative features, deciding on the number of speakers, and detection of overlapped speaker segments \cite{xavier12-review,king13-challenges}. A good segment initialization and speaker discriminative features help to improve the performance of the diarization system.
In this paper, we address these two challenges: (i) better segment initialization for IB based system, and (ii) use of conversation-specific speaker discriminative features in the diarization process. 

In the conventional IB based system, the initial speaker segments are obtained by dividing a conversation into short segments of equal duration.
{Each segment is modeled as a Gaussian and then clustered using IB criterion.}
It is essential to start with a good segmentation to obtain a good clustering solution.
In the first part of the paper, we propose a varying length segment initialization technique for the IB based system.
Initial segmentation ensures that the phoneme rate is fixed in each segment rather than the duration.
The phoneme rate is used as the side information in segment initialization such that the number of phonemes in each segment is approximately constant. 
This ensures that the boundaries of the segments are at the phone boundaries.
The proposed IB based system using the varying length segment initialization is referred to as  VarIB system.

Speaker diarization is a clustering task that involves grouping the speaker segments into different clusters.
The features used for the representation of speakers must be discriminative.
In the second part of this work, we focus on obtaining speaker discriminative features to improve the output of the IB clustering.
In~{\cite{dawalatabad16-tisd}}, we proposed the Two-Pass Information Bottleneck (TPIB) based speaker diarization using Multi-layer Feed-forward Neural Network (MFNN).
The TPIB system refines the speaker boundaries by performing the IB clustering process twice.
The speaker discriminative information from the output of the IB system is used to train MFNN based speaker classifier and obtain speaker discriminative features.
These features are then used again in the IB system to refine the speaker boundaries.

By design, the RTF of the TPIB system is at least twice the IB system.
Training MFNN is inherently an iterative process. 
To reduce the RTF of the TPIB system, one has to play with the hyperparameters of MFNN (e.g., the number of hidden layers, hidden nodes, etc.).
In this work, we propose a computationally simpler alternative to use Linear Discriminant Analysis (LDA)  {\cite{bishop}}.
LDA model can be trained significantly faster than MFNN, thereby reducing the overall runtime of the TPIB system.
We also propose a new cluster stopping criterion for LDA based systems.
From here onwards, we refer to TPIB systems with MFNN and with LDA as TPIB-NN and TPIB-LDA, respectively.
Finally, we propose a TPIB system with the varying length segment initialization using both MFNN and LDA.
This system is referred to as the VarTPIB system.

The aim of this work is to propose different architectures for information bottleneck based speaker diarization.
The following are the contributions to the paper.
\begin{enumerate}[leftmargin=*]
\item \textit{Segment initialization}: A varying length segment initialization (VarIB) technique using phoneme rate for information bottleneck based speaker diarization is proposed.
\item \textit{Speaker discrimination}: The TPIB-NN system was proposed in {\cite{dawalatabad16-tisd}}. TPIB-LDA system is proposed in this paper as an alternative to TPIB-NN to reduce the RTF. A novel stopping criterion for the TPIB-LDA system is also proposed. 
We also conduct an experimental study on the combination of latent features obtained using LDA and MFNN.
\item \textit{Joint segment initialization and speaker discrimination}: We propose VarTPIB approach, a combination of VarIB and TPIB approaches. It leverages both the benefits, i.e., good segment initialization and better speaker discrimination.
\end{enumerate}
The first two approaches focus on two different ways to improve the speaker diarization system, whereas the third approach jointly leverages the advantages of the first two proposed approaches.

The rest of the paper is organized as follows.
Section \ref{sec:prior} presents a review of related approaches to diarization.
Section \ref{sec:ib} describes the baseline IB based speaker diarization system.
Varying length segment initialization is proposed in Section  \ref{sec:varib}. 
Section \ref{sec:tpib} describes the TPIB and VarTPIB based systems.   
Different variants of the TPIB and VarTPIB systems are described in this section.
Results of experimental studies and analyses of results are presented in Section \ref{sec:exp}. 
Finally, Section \ref{sec:con} concludes the paper.


\section{Approaches to Speaker Diarization}
\label{sec:prior}

In this section, we discuss different approaches to speaker diarization reported in the literature related to the proposed approaches in this paper.

The performance of speaker diarization systems is dependent on the initialization of the segments to be clustered. 
In \cite{active-hansen}, the focus is to have good initial cluster models, leading to a significant reduction in the diarization error rate with minimal human supervision.
The approaches in \cite{Park2018,Park2019} use lexical information along with acoustic cues to improve the speaker diarization performance.
It is shown that the lexical information helps to add speaker discriminative information.
This approach employs a sequence-to-sequence neural network trained on both lexical and acoustic features.
Recently, we proposed a varying length segment initialization technique using the stroke onset rate information to diarize percussion instruments in Carnatic music concerts \cite{Dawalatabad2018}. It is shown that segment length initialization based on stroke rate improves the performance of the percussion instrument diarization system.

The accuracy of a speaker diarization system is also dependent on the feature representations of speakers.
A set of features that better discriminates one speaker from another speaker is preferred for diarization. 
Short term spectral features such as Mel Frequency Cepstral Coefficients (MFCC) are widely used for the task of speaker diarization \cite{xavier12-review}. 
The Mel Filterbank Slope (MFS) features and Linear Filterbank Slope (LFS) features have been shown to be better at speaker discrimination in comparison to MFCC features, owing to the emphasis of higher order formants \cite{srikanth14-mfs}. 
Group delay based features have also shown to be better at speaker discrimination over traditional MFCC \cite{padmanaban-gd}.
The i-vector based approaches have been used to build better speaker models  \cite{reynolds11-ivec,srikanth15-sgmm}. 
Linear Discriminant Analysis (LDA) has been used for speaker diarization of telephone data in \cite{lapidot12-lda}. 
Iterative probabilistic LDA adaptation for speaker diarization has been used for French TV recordings in \cite{le2016iterative}.

\begin{figure*}[t]
\centering
\includegraphics[scale=0.33]{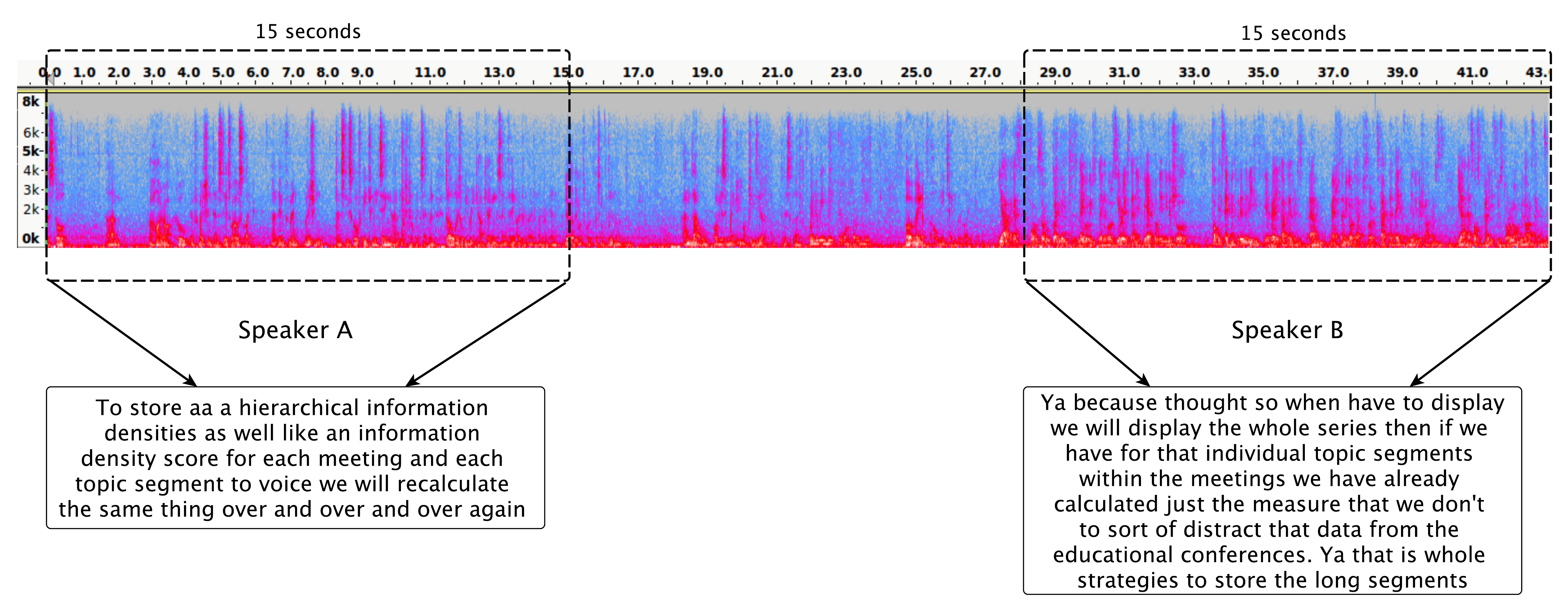}
\caption{Illustration of different speaking rates for different speakers. In a window of 15 seconds, speaker A spoke 35 words whereas speaker B spoke around 56 words. }
\label{fig:speaking_rate}
\end{figure*}

Recently, data-derived features extracted using  Deep Neural Networks (DNNs) are used.
In \cite{yella14-ann}, an Artificial Neural Network (ANN) is trained to determine whether a pair of utterances are from the same speaker or not. As the ANN is trained to be speaker discriminative, the features extracted from the ANN can be used to add discriminative feature information to the process of diarization.
The approach in \cite{le17-tripletloss} used the triplet ranking method for diarization and linking.
In \cite{Bredin2017TristouNetTL}, a DNN is used to get speaker embeddings. 
This approach is similar to \cite{yella14-ann}, but the Long Short Term Memory (LSTM) network trained with a triplet loss is used to get the embeddings.
In \cite{garcia17-dnnembedding}, a DNN is trained on large amounts of data to get speaker embeddings.
In \cite{sd-lstm}, an LSTM model is used to obtain speaker embeddings and perform spectral clustering on the same. 
In \cite{Lin2019}, a bidirectional LSTM is used to measure the similarity between speech segments.
The approach in \cite{sincnet-vtl} uses SincNet \cite{sincnet} in a Vanilla Transfer Learning (VTL) setup to obtain speaker discriminative embeddings.
End-to-end diarization techniques \cite{shinji19-e2e,fujita-19-e2e} have also been proposed where a single neural network directly outputs the diarized speaker boundaries.
DNN based embeddings such as x-vectors \cite{sell-dihard2018}, d-vectors \cite{dvector-1}, and c-vectors \cite{cvector}  are speaker discriminative.
In \cite{sup-sd}, a fully-supervised speaker diarization system using Unbounded-Interleaved State Recurrent Neural Network (UIS-RNN) is used in the clustering stage.
This technique assumes that high quality time-stamped speaker labels are available for training the UIS-RNN.
However, in all cases, the DNNs need to be trained on huge amounts of labeled data.

An incremental transfer learning based approach was proposed recently by us, which uses the ``Remember-Learn-Transfer" approach to continuously learn speaker discriminative information \cite{dawalatabad19:itl}.
This reduces the RTF of TPIB (with MFNN) system \cite{dawalatabad16-tisd}  by 33\% (relative) but with minor degradation in the speaker error rate.
It uses transfer learning \cite{Olivas:2009,Goodfellow:2016}  to retain the previously learned parameters, and adapts the parameters for the current conversation. 
In \cite{Jati2017}, attempts have been made to use {\it ``Speaker2Vec"} embeddings, where the system is trained on unlabeled data.
They have also suggested Neural Predictive Coding (NPC) \cite{jati2018:npc} where a Convolutional Neural Network (CNN) is trained in an unsupervised manner on separate data.
Despite several advancements in approaches to diarization, most of the recent approaches are based on deep embedding based techniques where embeddings are obtained in a supervised fashion, which requires a huge amount of speaker labeled data (thousands of hours \cite{xvector}) for training.

\section{Information Bottleneck based Approach}
\label{sec:ib}

In this section, we briefly describe the agglomerative Information Bottleneck (aIB) based approach to speaker diarization \cite{deepu09-ib}.

The agglomerative information bottleneck based approach for speaker diarization is based on the information bottleneck principle \cite{tishby2000-ib}.
An audio conversation is divided into fixed duration segments. 
Let $\mathbf{X}$ be a random variable representing a set of segments in the conversation.
Let $\mathbf{Y}$ be a random variable denoting a set of relevance variables that specify relevant information about the speaker in each segment. 
Let $\mathbf{C}$ be a clustering solution to $\mathbf{X}$.
Each segment is assumed to contain only one speaker and is modeled by a Gaussian distribution.
In the Gaussian Mixture Model (GMM) based modeling, the speaker information is represented by the components of the GMM.
Thus, the relevance variable $\mathbf{Y}$ represents the components of the GMM trained on the speech segments of an audio conversation.
Bottom-up clustering is performed in the posterior space of the Gaussian components.
The IB based approach clusters a set of segments $\mathbf{X}$ into a set of clusters $\mathbf{C}$ such that it captures as much relevant information as possible about $\mathbf{Y}$.
Thus, the objective function $\mathcal{F}$ is given by
\begin{equation}
\label{eq:obj}
\mathcal{F} =  \mathrm{I}( \mathbf{Y}, \mathbf{C}) - \frac{1}{\mathbf{\beta}} \mathrm{I}( \mathbf{C}, \mathbf{X})
\end{equation}

\noindent where $ \mathrm{I}(.)$ denotes the mutual information and \( \mathbf{\beta} \) is a Lagrange multiplier. 
The mutual information preserved $\mathrm{I}( \mathbf{Y}, \mathbf{C})$ is maximized while the mutual information $\mathrm{I}( \mathbf{C}, \mathbf{X})$ is minimized.

The Normalized Mutual Information (NMI) given by \( \frac{\mathrm{I}(\mathbf{Y,C})}{\mathrm{I}(\mathbf{X,Y})} \) is used as the stopping criterion \cite{deepu09-ib}.
Once clustering terminates, the output is resegmented using Kullback-Leibler Hidden Markov Model (KL-HMM) \cite{deepu09-klRealign}.
The IB based diarization is used as the baseline benchmark for comparison of the work reported in this paper.


\section{Varying Length Segment Initialization for IB (VarIB) based Diarization}
\label{sec:varib}

The first step in IB based diarization \cite{deepu09-ib} is the segmentation of an audio conversation into short segments of equal size.
Posteriors estimated from these segments are then clustered using IB based approach, as explained in Section \ref{sec:ib}.
The segment length does not change throughout the clustering process. Thus, the segments must be appropriately initialized in the IB based system.
Uniform segment initialization is not always an optimal strategy.
For instance, it can be seen from Fig. \ref{fig:speaking_rate} that in 15 seconds speaker A spoke around 35 words, whereas speaker B spoke around 56 words in the same amount of time.
 The speaking rate can vary significantly across different speakers.
The fixed length segment initialization may lead to a non-uniform (unequal) number of phonemes across segments.
This may lead to the non-uniform presence of speaker information across initial segments.
As a consequence, the parameters of the Gaussian model and the estimates of posteriors for clusters with fewer phonemes can be poor.

Motivated from \cite{active-hansen}, \cite{Park2018}, and \cite{Dawalatabad2018},  we propose a varying length segment initialization technique to address this problem.
 \cite{active-hansen} focuses on good initialization of clusters.
 In  \cite{Park2018}, it is shown that the lexical information adds speaker discriminative information to the diarization process.
In \cite{kumar2015musical}, group-delay processing is used to obtain the onset boundaries of strokes in a percussive instrumental music recording.
The information about the instrument is concentrated in the strokes.
In \cite{Dawalatabad2018}, the stroke rate is used as side information to initialize the segments for diarization such that the number of strokes (information about the instrument) across segments is approximately the same.

\begin{figure}[t]
\centering
\includegraphics[scale=0.21]{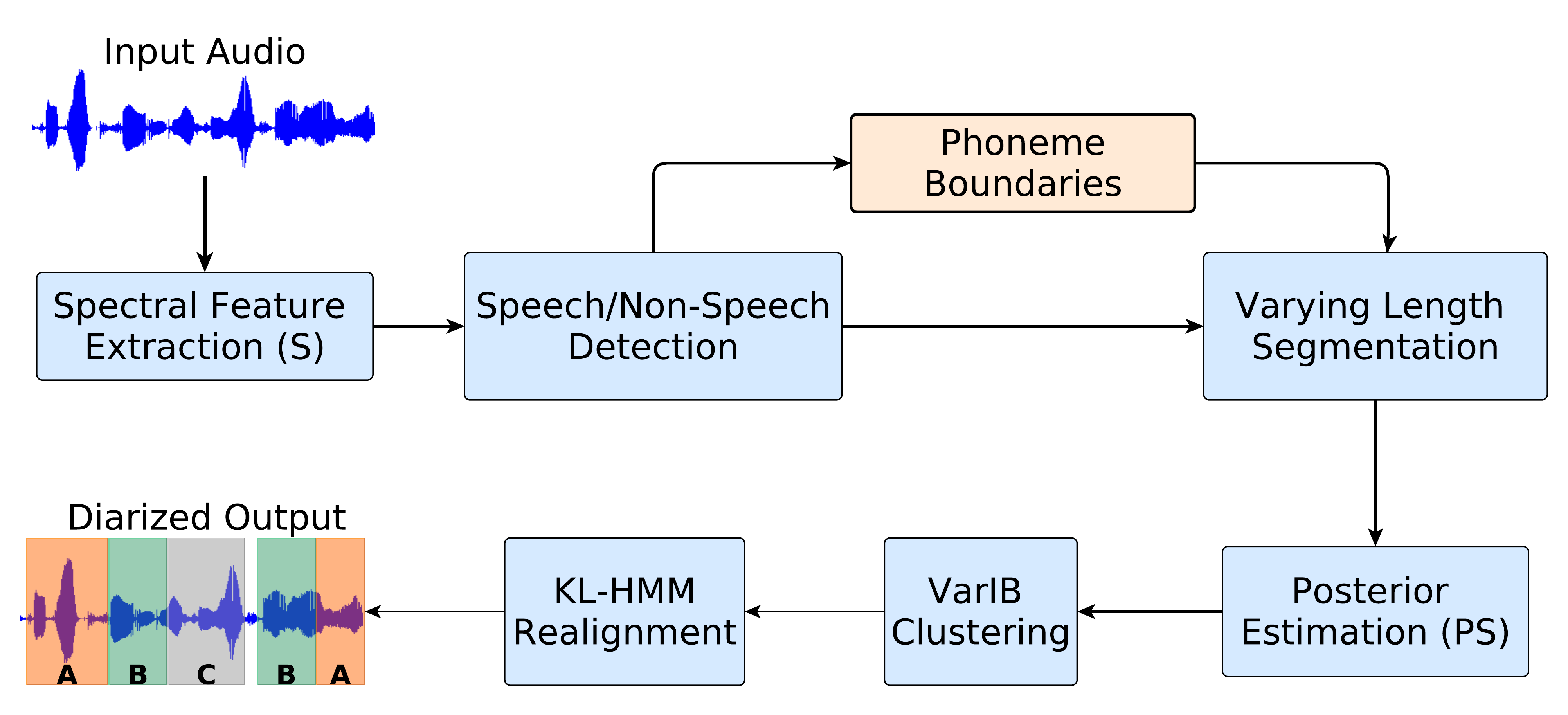}
\caption{Block diagram of VarIB based speaker diarization system.}
\label{fig:varIB_block}
\end{figure}

The speaker's information is present in the phonemes.
Phoneme rate is used to perform varying length segment initialization such that the number of phonemes is approximately the same across segments.
Phoneme boundaries are obtained using \textit{PhnRec} phoneme recognizer \cite{phnrec}. 
A range of durations in the interval [$minLen$, $maxLen$] is permitted. The chosen duration approximately corresponds to a fixed number of phonemes, as defined by $phnRate$, a hyperparameter. 
Here, $minLen$ and $maxLen$ are the minimum and maximum lengths of segments allowed, respectively.
The objective function $\mathcal{F}_{var}$ is given by

\begin{equation}
\label{eq:var_obj}
\mathcal{F}_{var} =  \mathrm{I}( \mathbf{Y}, \mathbf{C}) - \frac{1}{\mathbf{\beta}} \mathrm{I}( \mathbf{C}, \mathbf{X_{var}})
\end{equation}
where $\mathbf{X_{var}}$ represents the varying length segments.
The complete block diagram of the proposed VarIB system is shown in Fig. \ref{fig:varIB_block}.

The length of each segment is bounded between $minLen$ and $maxLen$ such that it has the $phnRate$ number of phonemes.
The number of varying length segments is dependent on these three hyperparameters.
These hyperparameters ensure that the number of segments is neither too small nor too large. 
Since the speaker's information is present in the set of phonemes he/she speaks, distributing an equal number of phonemes across the segments may lead to the same amount of information in each segment in a recording.
Since one Gaussian is used to model a segment, the proposed initialization approach ensures that a similar amount of information is used in modeling a Gaussian component from every segment.
Hence, the Gaussian models and the posteriors obtained using these models are expected to be better than those in the fixed length segment initialization in the IB based system.

\begin{algorithm}[t]
\vspace*{0.1cm}
\textbf{\underline{Input}}:\\ 
\vspace*{0.1cm}
\textit{Phoneme boundaries}, $minLen$, $maxLen$, and $phnRate$ \\
\vspace*{0.1cm}
\textbf{\underline{Output}}: \\ 
\vspace*{0.1cm}
 $X_{var}$: \textit{List of varying length segments} \\
\vspace*{0.1cm}
\textbf{\underline{Main Loop}}:\\ 
\vspace*{0.1cm}
  $ptr$ = 0 \\
 \While{  $ ptr <  len(audio)$  }
 {
$start$ = $ptr$ \\
$end$ = $ptr$ + $minLen$ \\
\If{$end \geq len(audio)$}
{
	$\{start,len(audio)\}$ $\Rightarrow$ $X_{var}$ \\
	$break$ \hspace{1cm} // \textit{for the final segment} \\
}  
$list_{A}$ = $getPhnList(start, end)$ \\

\eIf{$len(list_{A}) \geq phnRate$ }
{
	$\{start,end\}$ $\Rightarrow$ $X_{var}$   \\
	$ptr$ = $end$ \\
}
{ $list_{B}$ = $getPhnList(end, start + maxLen)$ \\
	
	\eIf{($len(list_{A}) +len(list_{B})) \leq phnRate$ }
	{
	$end$ = $list_{B} (len (list_{B}))$ \\
	$\{start,end\}$ $\Rightarrow$ $X_{var}$   \\
	$ptr$ = $end$ \\
	}
	{
	$\Delta{phnRate}$ = $phnRate$ - $len(list_{A})$ \\
	$end$ = $list_{B}(\Delta{phnRate})$  \\
	$\{start,end\}$ $\Rightarrow$ $X_{var}$   \\
	$ptr$ = $end$ \\
	}	
}
}
 \caption{Varying Length Segment Initialization}
\label{alg:var}
\end{algorithm}

The algorithm for varying length segment initialization is given in Algorithm \ref{alg:var}.
Here, $ptr$ is a pointer that points to the current time instant in the audio.
The function $getPhnList(t_1,t_2)$ returns the start and end times (boundaries) of phonemes between time instants $t_1$ and $t_2$.  
Given phoneme boundaries, $minLen$, $maxLen$, and $phnRate$; the algorithm returns the boundaries of varying length segments.
It may be noted that only phoneme boundaries are used.

\begin{figure*}[ht]
    \centering
    \begin{subfigure}[b]{0.45\textwidth}
    \centering
         \includegraphics[width=0.8\textwidth]{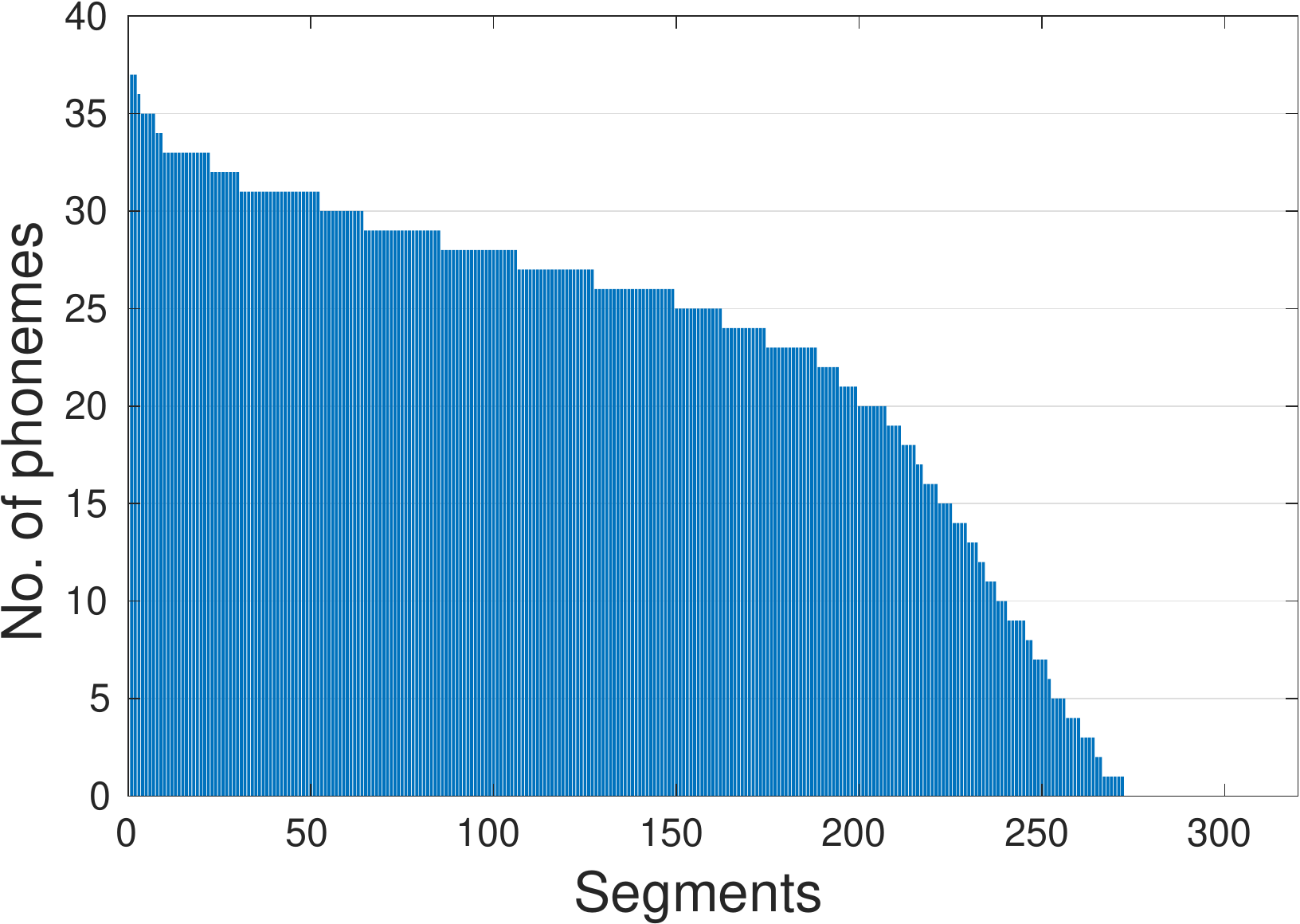}
        \caption{No. of phonemes in fixed length segments}
        \label{fig:fixed_phn_num}
    \end{subfigure}
    ~~~~~~
    \begin{subfigure}[b]{0.45\textwidth}
   \centering
       \includegraphics[width=0.84\textwidth]{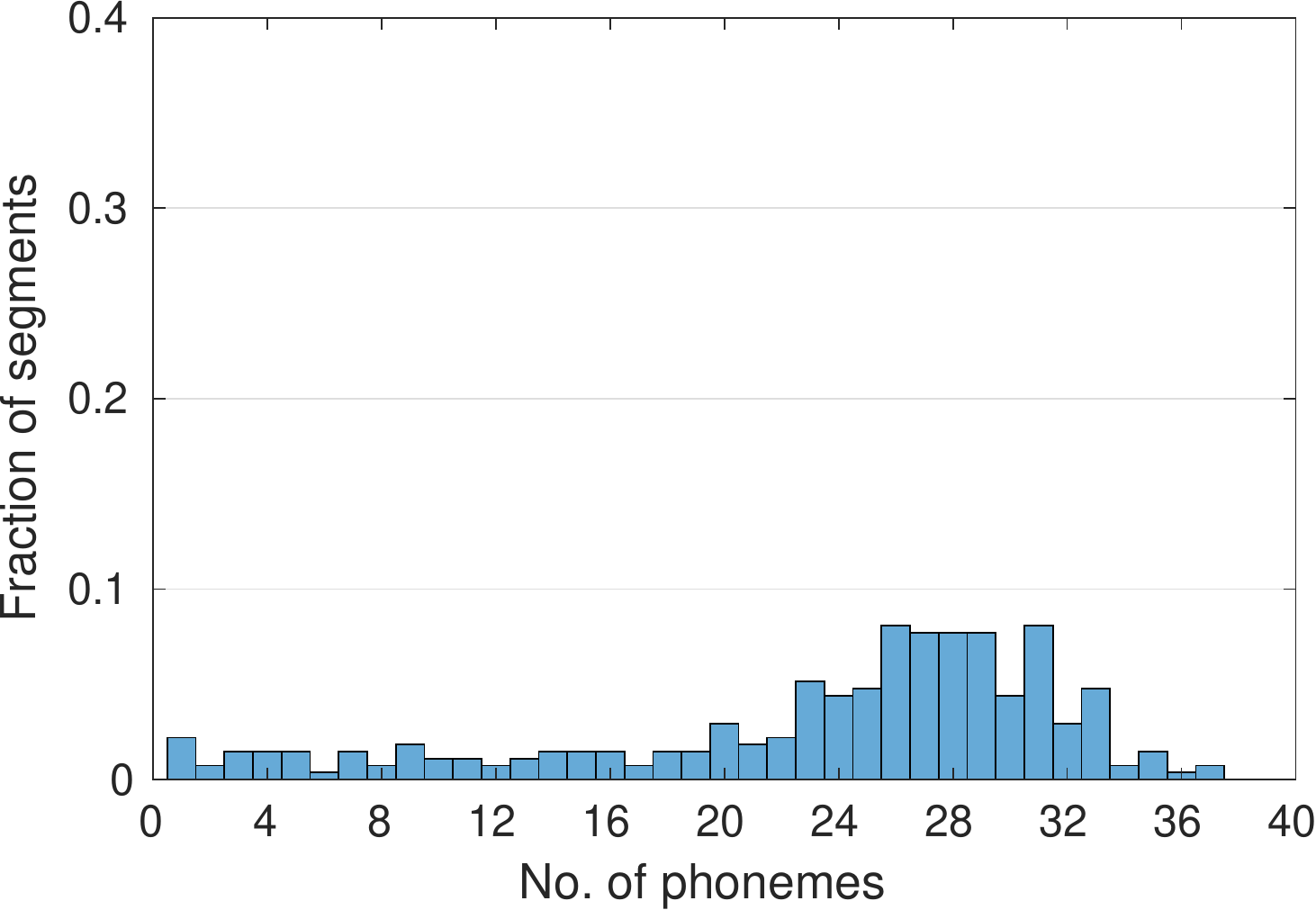}
        \caption{Distribution for fixed length segments in IB}
        \label{fig:fixed_phn}
    \end{subfigure}
        
\vspace{0.5cm}

    \begin{subfigure}[b]{0.45\textwidth}
     \centering
       \includegraphics[width=0.8\textwidth]{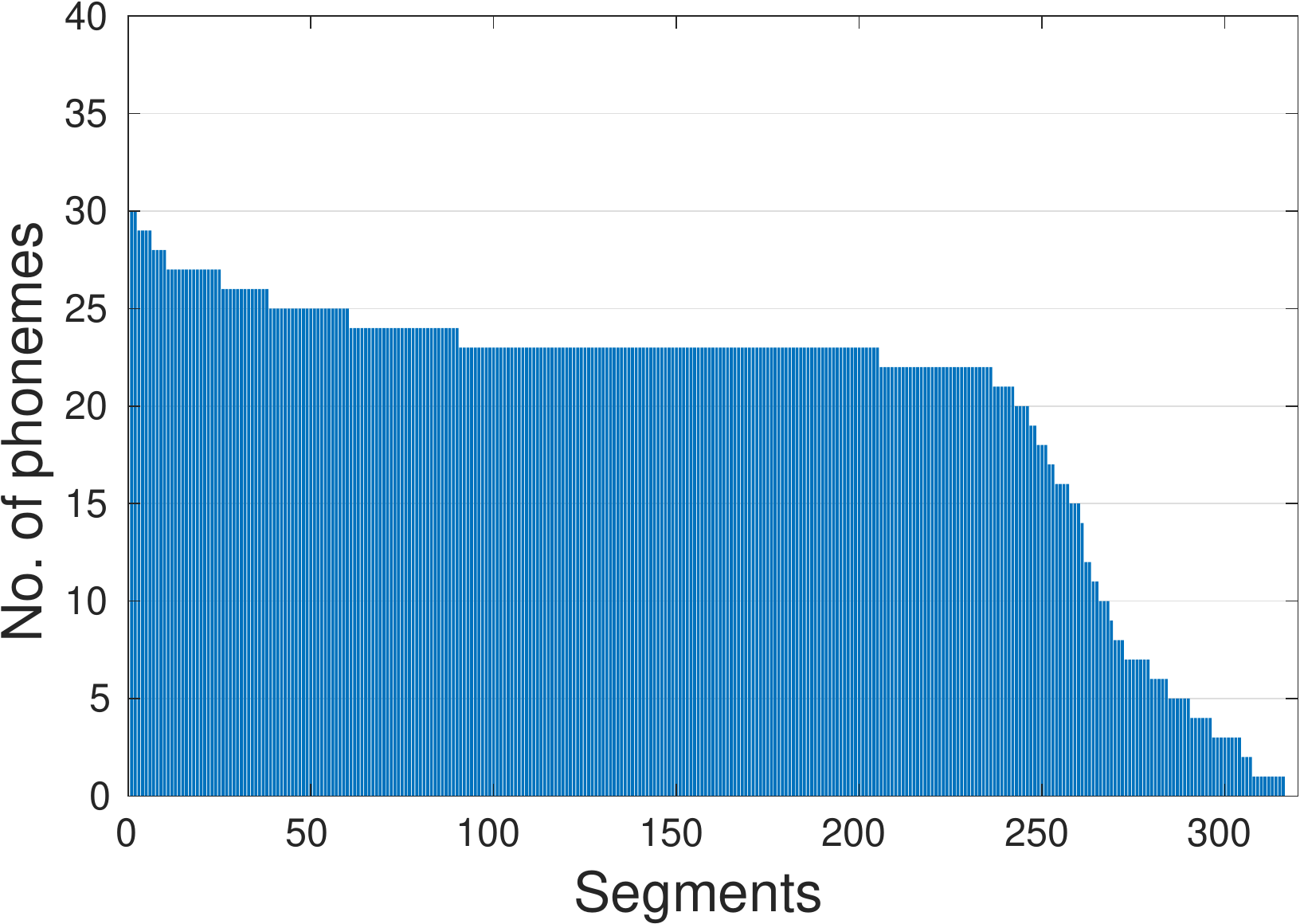}
        \caption{No. of phonemes in varying length segments}
        \label{fig:var_phn_num}
    \end{subfigure}
    ~~~~~~
    \begin{subfigure}[b]{0.45\textwidth}
    \centering
       \includegraphics[width=0.83\textwidth]{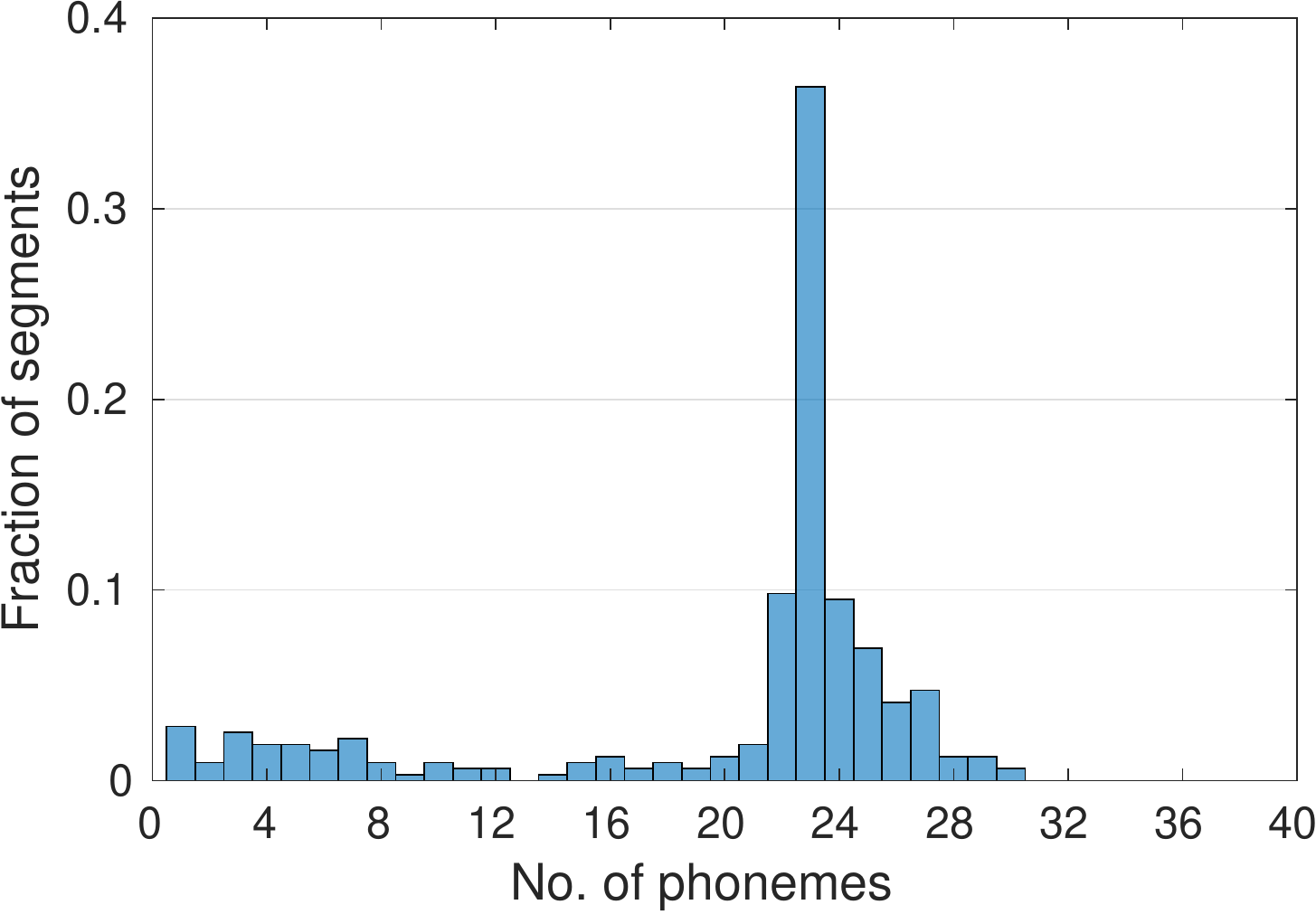}
        \caption{Distribution for varying length segments in VarIB}
        \label{fig:var_phn}
    \end{subfigure}
    
      \caption{Distribution of information across segments for fixed length segment initialization and varying length segment initialization.}
    \label{fig:phn}
\end{figure*}


\subsection{Posterior Estimation using Varying Length Segments}
\label{subsec:post_varIB}
A Gaussian component is modeled for each of the varying length segments.
The components collectively form a GMM.
The frame-level posteriors are obtained as given below.
\begin{equation}
	\label{eq:postvar}
	P(y_i|f_k) = \frac{a_i\mathcal{N}(f_k, \mu_i, \Sigma_i)}{\sum_{j=1}^{N} {a_j\mathcal{N}(f_k, \mu_j, \Sigma_j)}}
\end{equation}

\noindent where $f_k$ denotes a feature vector of a frame. 
Here,   $\mu_i$, $\Sigma_i$, and $a_{i}$ are the mean vector, covariance matrix, and the mixture coefficient for the $i$-th component in the GMM, respectively. $a_{i}$ is given by, 
\begin{equation}
	\label{eq:wtvar}
	a_{i} = \frac{T_{i}}{  \sum_{j=1}^{N} {T_{j}}  }
\end{equation}
where $T_{i}$ is the duration of the $i$-th segment in seconds, and $N$ is the total number of varying length segments to be clustered.
Segment level posteriors are obtained by averaging the frame-level posteriors for the frames in the segment. 	
The segment level posteriors are then used in IB based clustering, followed by one step of KL-HMM based realignment to adjust the speaker boundaries.


\subsection{Information Distribution Across Segments formed using Fixed and Varying Duration Segment Initialization}
\label{subsec:phn}

Figures \ref{fig:phn}(a) and \ref{fig:phn}(c) show the number of phonemes in different segments for fixed length (2.5 sec) segment initialization and for varying length segment initialization ($phnRate$=23) on a sample meeting recording  taken from the development dataset, respectively.
Segments are sorted in descending order of the number of phonemes in a segment.
The difference in the number of segments for the two cases is due to the respective segment initialization method.
It can be seen that the number of phonemes is distributed unevenly in the fixed duration segment initialization. In contrast, in the varying length segment initialization, most of the segments have approximately the same number of phonemes.

Figures \ref{fig:phn}(b) and \ref{fig:phn}(d) correspond to the distributions of segments based on number of phonemes for Figures \ref{fig:phn}(a) and \ref{fig:phn}(c), respectively.
Long tails on the left and right correspond to low speaking rates and high speaking rates, respectively.
It can be seen that the distribution in Fig. \ref{fig:phn}(d) is peaky (less variance) compared to the distribution in Fig. \ref{fig:phn}(b).
The distribution in Fig. \ref{fig:phn}(d) peaks at 23 number of phonemes, indicating that most segments comprise of 23 (here, $phnRate$) phonemes.
The region to the immediate left of the peak represents segments from slow ($< phnRate$) speaking parts, whereas the region right to the peak shows short segments in fast ($> phnRate$) speaking regions.
The segments in the faster speaking regions are set to $minLen$ duration, whereas other segments are of lengths between $minLen$ and $maxLen$.

It should be noted that both the distributions include short speech segments of duration less than $minLen$ in the proposed system and segments less than the fixed duration in the baseline IB system.
These segments are the residual segments that result when big chunks of audio are divided into short uniform or varying length segments, as can be seen from Figures \ref{fig:phn}(b) and \ref{fig:phn}(d), respectively.
The distribution of residual segments is observed in the extreme left tail of the distributions.
However, the number of such short residual segments is small and is less than (or equal to) the number of speech regions in a conversation audio.

\begin{figure*}[t]
\centering
\includegraphics[scale=0.23]{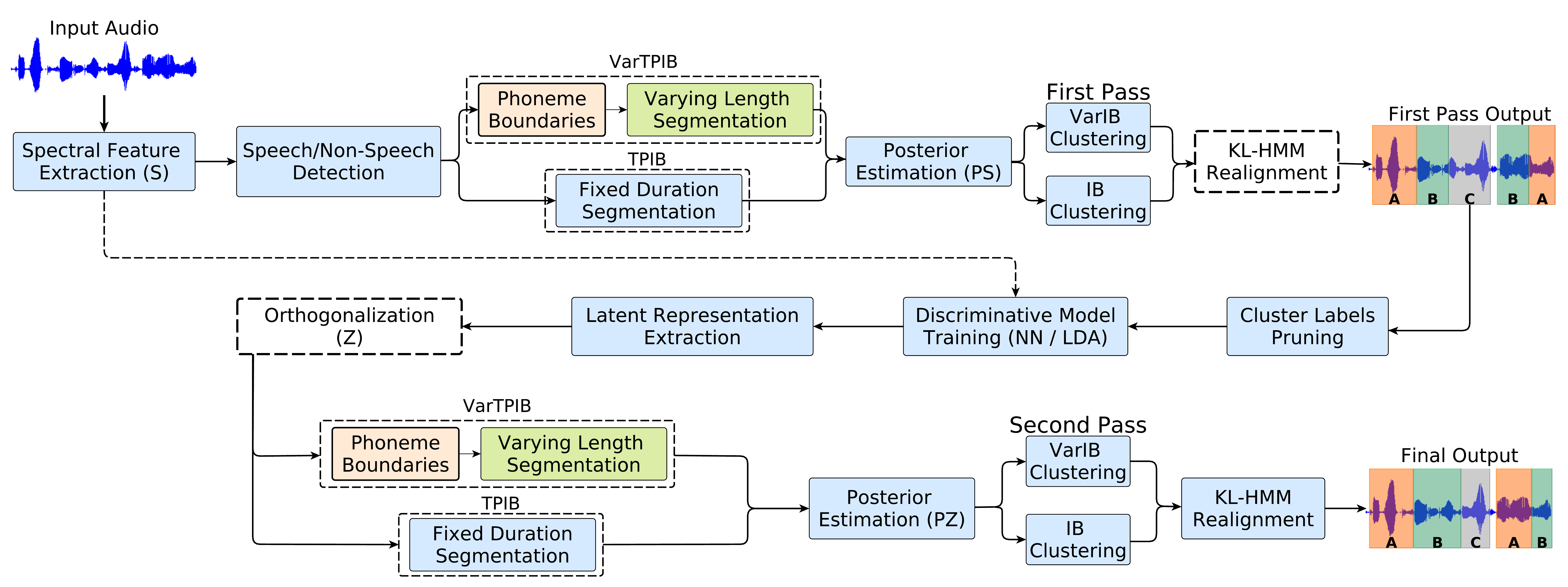}
\caption{Two-Pass IB (TPIB) and Varying Length Two-Pass IB (VarTPIB) based frameworks for speaker diarization. The KL-HMM based realignment after the first pass and the orthogonalization are performed only when a neural network model is used as the speaker discriminative model. 
The VarTPIB system uses phoneme rate as side information for segment initialization, while the TPIB framework uses a fixed duration segment initialization.}
\label{fig:tpib_vartpib}
\end{figure*}


\section{Two-Pass IB (TPIB) and Varying Length Two-Pass IB (VarTPIB) based Framework}
\label{sec:tpib}

In this section, we propose TPIB and VarTPIB approaches to include speaker discriminative features within the IB clustering framework.
This section first describes the general Two-Pass IB (TPIB) and Varying Length Two-Pass IB (VarTPIB) based frameworks for speaker diarization, as shown in Fig. \ref{fig:tpib_vartpib}.
We then describe different discriminative models (MFNN and LDA) used in this framework. 

The TPIB-LDA approach presented in this section is an extension to our previous work \cite{dawalatabad16-tisd} where the TPIB-NN (as discussed later in this section) system was first introduced.
The VarTPIB approaches proposed in this section is the combination of VarIB and TPIB approaches. 
In general, the TPIB and VarTPIB framework consists of the following stages.

\begin{itemize}[leftmargin=*]
\item \textit{Spectral Feature Extraction:}
    This stage takes a conversational audio as input and extracts acoustic features.
    Speech/non-speech segments can be identified using a Speech Activity Detection (SAD) {{\cite{rabiner}}}.
    
\item \textit{Segment Initialization:}
    Input audio is divided into short segments of uniform duration in the TPIB framework.
    The segment length is usually fixed to 2.5 seconds \cite{deepu09-ib}.
 VarTPIB approach initializes segments based on phoneme rate information as given in Algorithm {\ref{alg:var}}.
    
\item \textit{Posterior Estimation:}
    Each of the $N$ segments in the TPIB approach is initially modeled by a single Gaussian.   
    These $N$ Gaussians are used to build a GMM. 
    The posterior for each frame of speech $f_{k}$ for the $i$-th GMM component is computed using Equation (\ref{eq:postvar}) with mixture coefficients set to the value of $1/N$.
    The posteriors are then averaged over the frames in the segment to obtain segment level posteriors. 
    Finally, each segment is represented by an $N$-dimensional posterior vector ($P_s$). 
     The frame level and the segment level posterior estimation for VarTPIB is the same as explained in Section \ref{subsec:post_varIB}.
        
\item \textit{First Pass:}
    The segment level posteriors obtained are then clustered in a bottom-up fashion using the IB (or VarIB) approach as described in Section \ref{sec:ib} (or \ref{sec:varib}).
    The clustering is stopped when NMI reaches a threshold value. 
    Once clustering is completed, the speaker boundaries are refined using the KL-HMM  \cite{deepu09-klRealign} based approach.

\item \textit{Discriminative Model Training:}
    This stage takes the spectral features and the cluster labels for a feature vector from the first pass as input.
   Assuming that the number of speakers obtained during the first pass is correct, a classifier is trained to discriminate between the speakers obtained in the first pass.
   Clusters smaller than a threshold duration ($<$ 3~sec) obtained in the first pass are pruned out and not considered in training the classifier.
    The discriminative model is chosen such that the time taken to train the model is small.
    Different discriminative models used in this work are discussed in Section \ref{subsec:model}.

\item \textit{{Latent Representation Extraction}:}      
The discriminative model trained in the previous step is used to project the spectral features onto a latent space where the features are discriminative.  
The projection onto the latent space varies based on the discriminative model used.

\item \textit{Orthogonalization:}
    The discriminative features are then subjected to Principal Component Analysis (PCA).    
    At the end of this process, the feature directions are orthogonal, leading to a diagonal covariance matrix, a requirement for IB based clustering\footnote{PCA is used only for orthogonalization, and not for dimension reduction.}.
    The features obtained after orthogonalization are referred to as the Latent Features (LF).
    The latent features form a stream like the spectral feature stream.

\item \textit{Posterior Estimation for Latent Feature Stream:}
   In the TPIB framework, the LF stream for conversation audio is first uniformly segmented.
   The posteriors are then obtained as given in Equation (\ref{eq:postvar}), where $a_i$ is set to $1/N$ for TPIB system.
The frame-level posteriors are then used to obtain segment level posteriors ($P_z$).
Segment level posteriors ($P_z$) for VarTPIB framework is calculated from the LF stream as mentioned in Section \ref{subsec:post_varIB}.

\item \textit{Second Pass:}
The IB (or VarIB) based clustering is performed on $P_z$ in the second pass.
Finally, one step of KL-HMM \cite{deepu09-klRealign} based realignment is performed to refine the speaker boundaries that forms the final output of the TPIB/VarTPIB framework.

\end{itemize}

\subsection{Speaker Discriminative Models in the TPIB and VarTPIB Frameworks}
\label{subsec:model}
In this section, we describe two methods used in the \textit{Discriminative Model Training} stage of TPIB/VarTPIB framework.

\subsubsection{Mutli-Layer Feed-Forward Neural Network (MFNN)}
\label{subsec:ann}

After obtaining the speaker boundaries from the first pass, an MFNN is trained to classify the speakers in a conversation. 
The inputs to the MFNN are the spectral features of the frames. 
The labels obtained from the first pass are used as the desired output.
To keep the RTF of the entire system low, a shallow MFNN with two hidden layers is used. 
Sigmoidal activation functions are used in the hidden nodes in the first hidden layer and  linear nodes are used in the second hidden layer.
The MFNN is first trained using gradient descent to classify speakers with cross-entropy loss function.
Once the MFNN is trained, the spectral features of a frame of speech are given as input to the MFNN to obtain corresponding latent representation from the second hidden layer.
The latent features obtained after orthogonalization are denoted as $LF_{NN}$.
The $LF_{NN}$ obtained from the trained MFNN are used as a feature stream during the second pass of IB clustering.
The remaining steps are identical to those presented in Section \ref{sec:tpib}.
The two pass IB/VarIB system  with discriminative features obtained using an MFNN is referred to as TPIB-NN/VarTPIB-NN.   
Although care has been taken in TPIB-NN/VarTPIB-NN to keep the number of model parameters small, the run-time of TPIB-NN/VarTPIB-NN system is comparatively high because of the back-propagation algorithm used to train the MFNN.
However, the RTF is still well below 1.0 (as mentioned later in Table \ref{tab:rtf}).

\subsubsection{Linear Discriminant Analysis (LDA)}
\label{subsec:lda}
In this section, we propose to reduce the RTF of TPIB-NN and VarTPIB-NN systems  by using a non-iterative discriminative method.
The LDA  can be performed in significantly less amount of time than the MFNN based method.
The TPIB/VarTPIB system using LDA as speaker discriminator is denoted as TPIB-LDA/VarTPIB-LDA.
The spectral feature extraction and the posterior estimation stages are identical to those discussed in  Section \ref{sec:tpib}. 
The following stages are different from that of TPIB-NN/VarTPIB-NN system (i.e., Section \ref{subsec:ann}).

In TPIB-LDA/VarTPIB-LDA, the IB clustering process in the first pass is stopped when the total number of clusters reaches a fixed number.
This is because the upper limit on the reduced dimension in the LDA is one less than the  number of classes (speaker clusters).
When the number of speakers is small, the LDA based dimension reduction can lead to a significant loss of information.
In the proposed system, LDA is used to obtain  discriminative latent features and not for dimension reduction.
The number of clusters is chosen empirically based on the performance on the development data. 
This is not required in case of TPIB-NN/VarTPIB-NN system as the number of nodes in the second hidden layer can be kept as high as the input spectral dimension itself.
Details of the stopping criterion for LDA based systems is discussed in Section \ref{subsec:stopping}.
The realignment can result in a significant reduction in the number of speaker clusters as smaller clusters may get assigned to a bigger cluster.
Thus, unlike TPIB-NN/VarTPIB-NN system, KL-HMM realignment is not performed after the first pass of IB clustering in TPIB-LDA/VarTPIB-LDA.
This is indicated by a dotted block in Fig. \ref{fig:tpib_vartpib}.

The cluster labels from the first pass along with the spectral feature stream are used, and LDA is applied.
We then extract the latent representation by projecting spectral features onto the LDA space. 
The obtained latent representation is referred to as $LF_{LDA}$.
The input spectral features are already independent, and the LDA performs a linear transformation which retains the independence among the features even in the latent space. 
Hence, latent representations are not subjected to orthogonalization.
The second pass of IB based clustering is performed on $LF_{LDA}$ with NMI as the stopping criterion.

\begin{figure}[]
\centering
\includegraphics[width=0.37\textwidth]{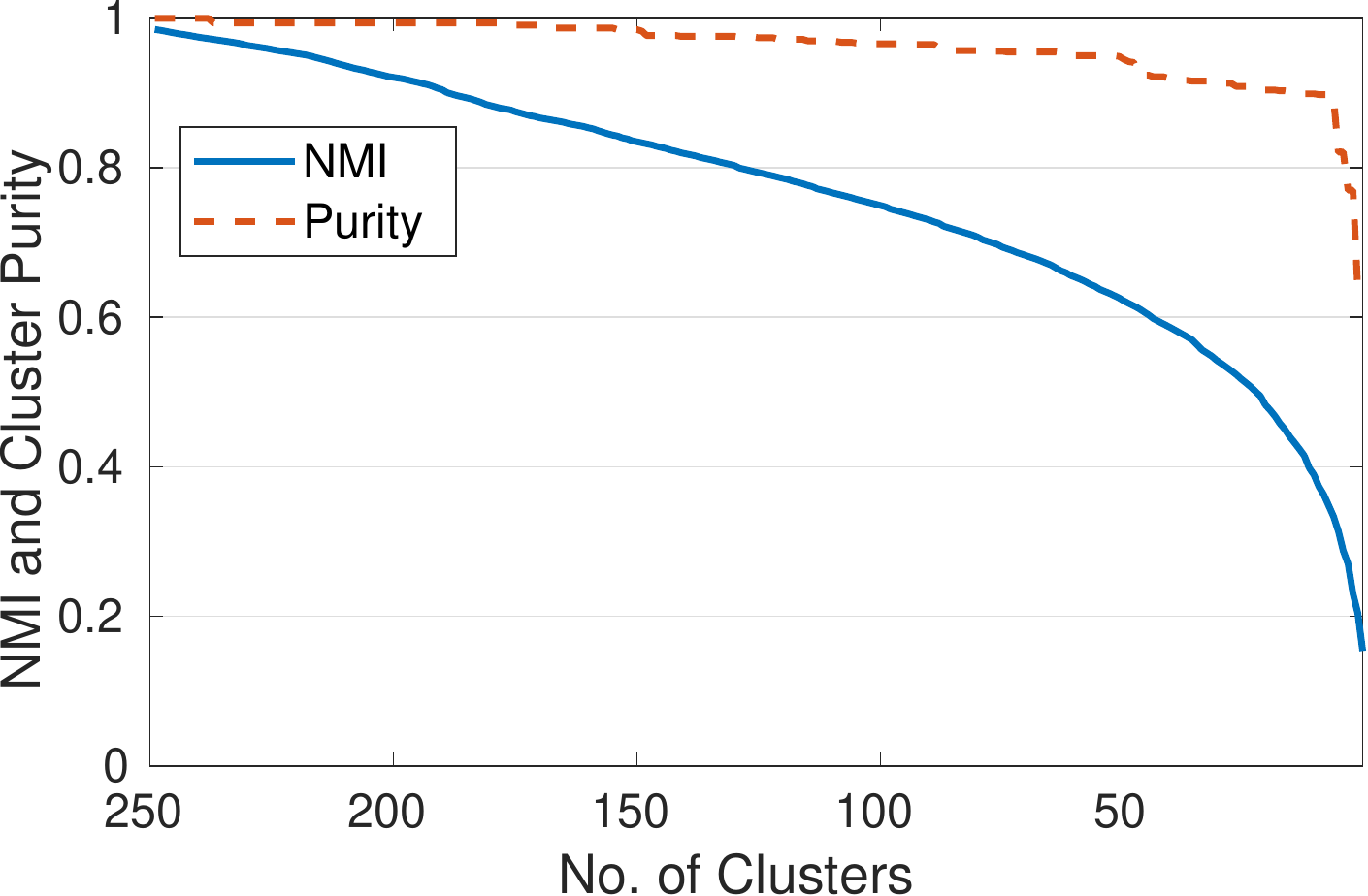}
\caption{NMI and cluster purity at different iterations (number of clusters) in IB based bottom-up clustering for a recording.}
\label{fig:nmi_pur}
\end{figure}

\begin{figure*}[t]
\centering
\includegraphics[width=0.95\textwidth]{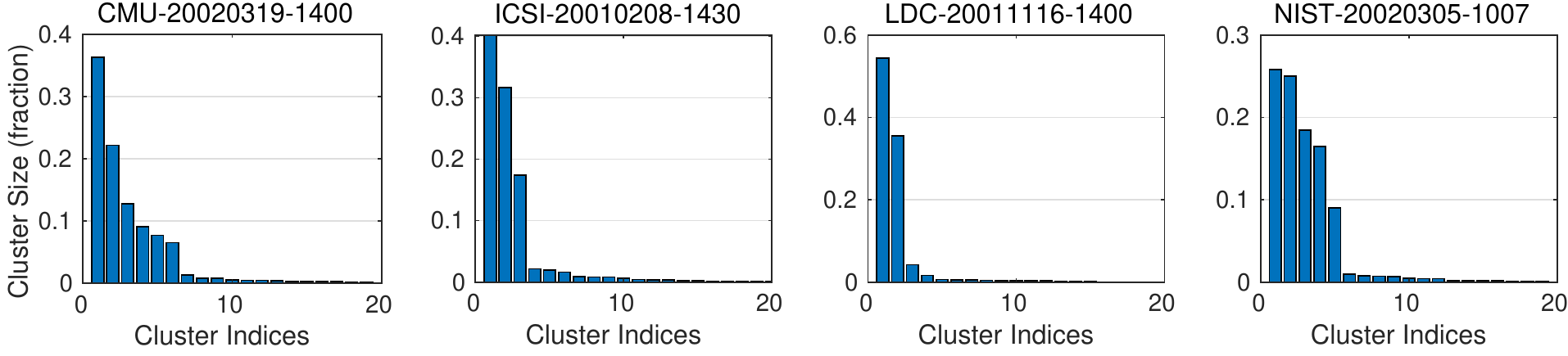}
\caption{Size of each cluster given as a fraction of the length of recording for 4 different recordings in development dataset for TPIB-LDA system. The clusters are sorted in the descending order of their sizes.}
\label{fig:clust}
\end{figure*}

\subsection{Stopping Criterion for TPIB-LDA and VarTPIB-LDA Systems} 
\label{subsec:stopping}

A proper stopping criterion in diarization is crucial as it  affects the final diarization error rate (DER) \cite{king13-challenges}.
Stopping the clustering process at the number of clusters less than the actual number of speakers is termed as over-clustering, and the opposite is referred to as under-clustering \cite{xavier12-review}. 
Under-clustering is preferred to over-clustering, as over-clustering can lead to merging of many speakers into a single cluster \cite{king13-challenges,comp_study}.
Fig. \ref{fig:nmi_pur} shows the NMI and cluster purity\footnote{Cluster purity is calculated using open source tool \textit{pyannote metric} \cite{pyannote.metrics}.} at different iterations in IB based bottom-up clustering.
It can be seen that as the number of clusters decreases, the NMI and cluster purity keeps on decreasing.
The clusters become impure (multiple speakers are in one cluster) as the bottom-up clustering proceeds.
This leads to an increase in the DER.

The number of speakers in a recording can be as small as two.
Unlike MFNN, there is less flexibility in LDA with respect to the choice of the number of dimensions, as the number of dimensions is decided by the number of classes.
The LDA reduces the dimension maximum to the \textit{number of classes - 1}.
For example, the LDA could reduce the MFCC feature dimension from 19 to 1 when there are only two speakers in a conversation.
This dimension reduction is highly lossy.
Hence, to avoid the loss of speaker information, we propose to stop the first pass of the TPIB-LDA and VarTPIB-LDA systems at a larger number of clusters (here, 20). 
This under-clustering in the first pass of IB clustering eliminates the problem of loss of speaker information.

Since  LDA performs linear discrimination among the speakers in the output of the first pass, it can separate similar speakers when the first pass is under-clustered.
Fig. \ref{fig:clust} shows the sizes of different speaker clusters sorted in descending order for 4 recordings taken from  the development set.
The plots show the cluster sizes when the number of clusters is 20.
It can be seen that all the recordings have  two or three dominating speakers.
The top four or five clusters contribute to more than 95\% of the speaker segments while the rest of the clusters have negligible size.
The small size clusters have small prior and hence do not contribute much when LDA is performed.

\subsection{Conversation-Specific Speaker Discriminative Features}
\label{subsec:disc}

Most of the approaches to find speaker discriminative features for speaker diarization task directly use techniques used in the speaker recognition domain.
The amount of training data required to define a global speaker discriminative space robustly can be quite large.
In the task of speaker recognition, it is necessary to find a global set of speaker discriminating features for a large number of speakers.
Finding a universal set of speaker discriminative features is a difficult problem. 
Further, the notion of the universal number of speakers is not well defined which makes the task more complicated.
On the other hand, in a speaker diarization task, the system is only required to discriminate between speakers in the given conversation recording.
The number of speakers in a conversational recording is very small (typically of the order of 2-10).
Therefore, only the relative discriminating characteristics need to be found.
Hence, finding  speaker discriminative features for speaker diarization task is relatively simple.

In the TPIB/VarTPIB architectures, the discriminator model is trained to discriminate the speakers in the current conversation recording.
Hence, the latent features obtained from the model are expected to be speaker discriminative specific to the current recording.
It is important to note that the proposed architectures do not use any external data (labeled or unlabeled) to train the models to obtain speaker discriminative features. 
Since discriminative models are trained on a recording that is to be diarized, the TPIB/VarTPIB systems do not suffer from an unseen class (speakers) problem.
Conversation-specific speaker discriminative features are obtained during the diarization process itself.

\section{Experimental Studies}
\label{sec:exp}
In this section, we demonstrate the effectiveness of the proposed architectures on standard meeting conversation datasets.

\begin{table}[t]
\centering
\caption{AMI meeting datasets.}
\label{tab:ami}
\resizebox{0.45\textwidth}{!}{
\begin{tabular}{@{}c|c@{}}
\toprule
\textbf{AMI-1} & \begin{tabular}[c]{@{}c@{}}ES2008c, ES2013a, ES2013c, ES2014d, ES2015a,\\ IS1001c, IS1007a, IS1008c, IS1008d, IS1009c\end{tabular} \\ \midrule
\textbf{AMI-2}          & \begin{tabular}[c]{@{}c@{}}ES2010b, ES2013b, ES2014c, ES2015b, ES2015c,\\ IS1004b, IS1006c, IS1007c, IS1008a, IS1009d\end{tabular}          \\ \bottomrule
\end{tabular}}
\end{table}

\subsection{Datasets, Features and Evaluation Measure}
\label{subsec:data}
\subsubsection*{Datasets}
Experiments are performed on the following standard NIST-RT datasets from Linguistic Data Consortium \cite{ldc}: RT-04Dev, RT-04Eval, and RT-05Eval. 
The subsets of  Augmented Multi-Party Interaction (AMI) corpus \cite{ami} are also used in our experiments.
All datasets contain recordings of conversational audio meetings.
The list of meeting IDs from AMI datasets recorded at Idiap (IS) and Edinburgh (ES) is given in Table \ref{tab:ami}.
The datasets are also used in \cite{dawalatabad16-tisd} and \cite{dawalatabad19:itl} for evaluation. 
The number of speakers in the NIST dataset varies from 3 to 10.
There are four speakers in each of the AMI recordings.
RT-04Dev is used as the development dataset to tune hyperparameters of all models, and the remaining datasets are used for testing. 
Multiple distant microphone (MDM) data is used for each meeting after beamforming using \textit{BeamformIt}~\cite{xavier07-beamformit}.

\subsubsection*{Features}
Since the focus of this work is on the different architectures and not on the feature side, we demonstrate the effectiveness of the proposed approaches on most widely used standard Mel Frequency Cepstral Coefficients (MFCC) features.
MFCC of 19 dimensions  extracted from 26 filterbanks\footnote{Overall SERs across all datasets on baseline IB system is better than that reported in~\cite{dawalatabad16-tisd} as we use 26 filterbanks in this paper as opposed to 40 filterbanks in~\cite{dawalatabad16-tisd}.}  are used as spectral features  \cite{deepu09-ib} \cite{deepu09-klRealign} \cite{srikanth14-mfs}. 
These MFCCs are obtained with 25ms window length and 10ms shift.

\subsubsection*{Evaluation Measure}
The Diarization Error Rate (DER) is the sum of Missed Speech (MS), False Alarm Speech (FS), and Speaker Error Rate (SER).
MS and FS depend on errors in speech activity detection, whereas SER arises due to speaker mismatch.
As the primary focus of this paper is on improving the clustering module, the speech/non-speech hypotheses are obtained from the ground truth.
Hence, similar to \cite{dawalatabad16-tisd,srikanth14-mfs}, we use SER as the evaluation metric. 
We used the standard diarization evaluation tool from NIST \cite{nist} to calculate SER with a forgiveness collar of 25ms.
Similar to \cite{deepu09-ib,dawalatabad16-tisd,srikanth14-mfs}, we also included the overlapped speech in evaluation in all our experiments.

\begin{table*}[]
\centering
\caption{Speaker Error Rates (SERs) for different systems. The best case performance for each dataset is given in bold font. The SER of all the proposed architectures is compared with the baseline IB system.}
\resizebox{\textwidth}{!}{
\begin{tabular}{ccccccccc}
\toprule
\multirow{2}{*}{System}     & \multirow{2}{*}{Segment Initialization} & \multirow{2}{*}{Discriminative Model(s)} & \multirow{2}{*}{Features} & Dev      & \multicolumn{4}{c}{Test Set}              \\
\cmidrule(l){6-9} \cmidrule(l){5-5}
                         &                               &                           &                           & RT-04Dev & RT-04Eval & RT-05Eval & AMI-1 & AMI-2 \\ \midrule
IB                       & Fixed                         & -                         & MFCC                      & 15.1     & 13.5      & 16.4      & 17.9  & 23.5  \\ \midrule 
\multicolumn{9}{c}{Proposed Systems}                                                                                                                                \\ \midrule 
VarIB                    & Varying                       & -                         & MFCC                      & 12.3     & 12        & 15.3      & 17.8  & 22.6  \\ \midrule
\multirow{4}{*}{TPIB}    & \multirow{4}{*}{Fixed}        & MFNN                       & $LF_{NN}$                   & 14.2     & 12.6      & 14.2      & 16.1  & 23.6  \\
                         &                               & LDA                       & $LF_{LDA}$                   & 14.7     & 11.6      & 13.2      & 15.7  & 24.5  \\
                         &                               & MFNN+LDA                   & $LF_{NN}+LF_{LDA}$ (0.2,0.8) & 13.1     & 12.6      & 12.6      & 15.4  & 21.9  \\
                         &                               & MFNN+LDA                   & $LF_{NN}+LF_{LDA}$ (Avg.)    & 14.2     & 12.4      & 14.5      & 16.3  & 22.2  \\ \midrule 
\multirow{4}{*}{VarTPIB} & \multirow{4}{*}{Varying}      & MFNN                       & $LF_{NN}$                   & 12       & \textbf{9.9}       & 14.2      & 17.5  & \textbf{20.9}  \\
                         &                               & LDA                       & $LF_{LDA}$                   & 13.8     & 12.8      & \textbf{12.5}      & 14.8  & 21.3  \\
                         &                               & MFNN+LDA                   & $LF_{NN}+LF_{LDA}$ (0.6,0.4)          & \textbf{11.6}     & 11.7      & 15.1      & \textbf{13.2}  & 21.1  \\
                         &                               & MFNN+LDA                   & $LF_{NN}+LF_{LDA}$ (Avg.)    & 12.6     & 11.9      & 13.9      & 15.1  & 21.1 \\
                         \bottomrule
\end{tabular}
}%
\label{tab:ser}
\end{table*}

\subsection{Experimental Setup}
\label{subsec:exp_setup}

We use the open source IB toolkit\footnote{Code is available at: https://github.com/idiap/IBDiarization}~\cite{deepu12:diartk} in  all our experiments.
The values for NMI and $\beta$ for all the systems for both the passes are set to 0.4 and 10, respectively.

\begin{figure}[]
\centering
\includegraphics[width=0.47\textwidth]{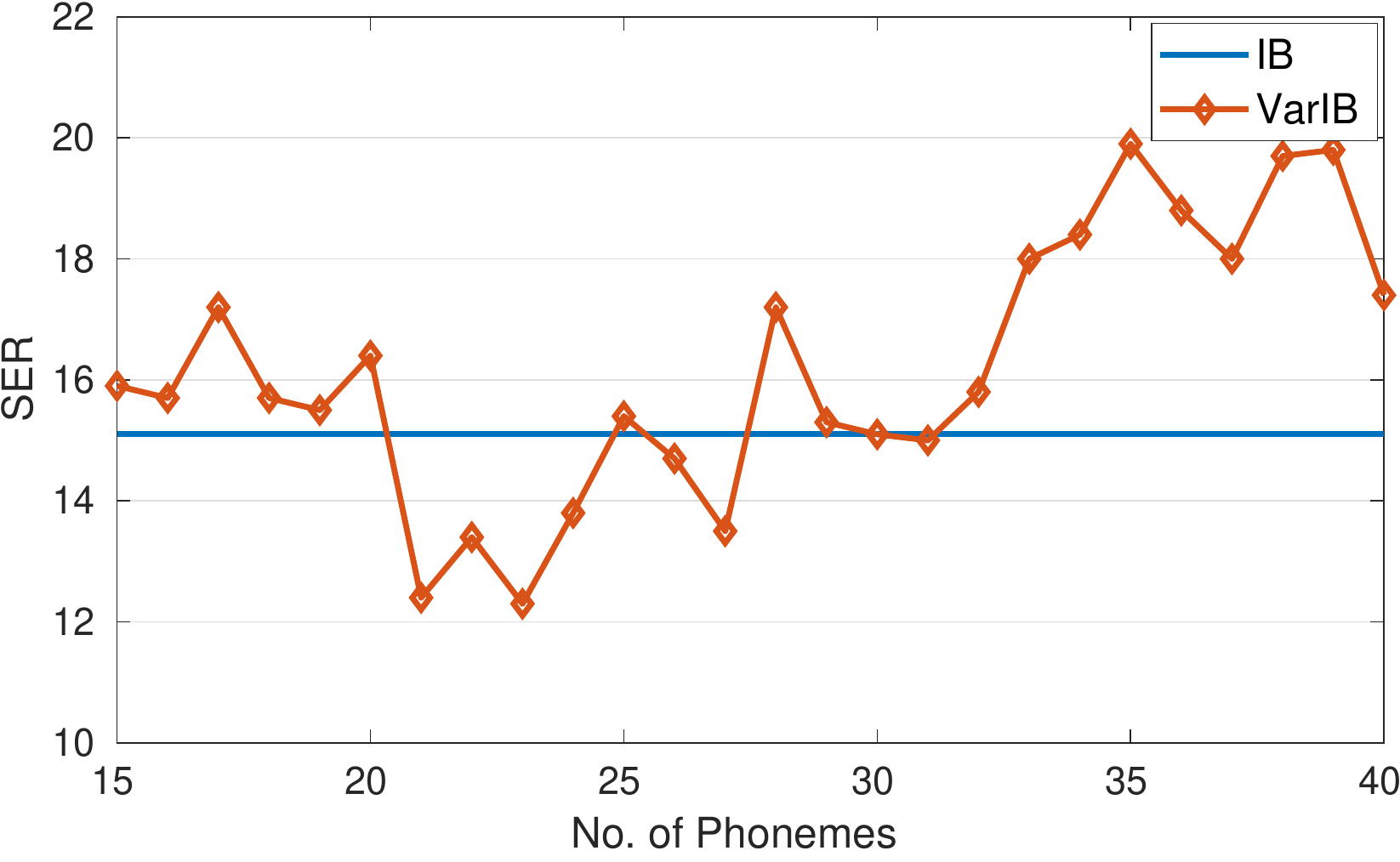}
\caption{SERs on development data as a function of window length in terms of number of phonemes (i.e. $phnRate$) in VarIB system. A reference line (in blue) denotes SER of IB based diarization system.}
\label{fig:varib_phnrate}
\end{figure}

\subsubsection*{VarIB system}
We use $PhnRec$\footnote{Code is available at: https://speech.fit.vutbr.cz/software/phoneme-recognizer-based-long-temporal-context} to detect the phoneme boundaries. 
Note that $PhnRec$ is used only to detect boundaries as the actual identities of the phonemes are not needed.
Fig. \ref{fig:varib_phnrate} shows SERs on development dataset for different number of phonemes ($phnRate$) in VarIB system.
The blue line is a reference line denoting SER on the development dataset (15.1\%) for the baseline IB based diarization system.
The $phnRate$ is varied between 15-40 phonemes, which corresponds to 4-12 syllables approximately. 
A reduction in SER is observed between 21-24 phonemes.   
The lowest SER is observed at 23 phonemes.   
The values of $minLen$ and $maxLen$ are set to 2 seconds and 5 seconds, respectively.
These hyper-parameters are kept the same across VarIB, VarTPIB-NN, and VarTPIB-LDA systems.

\subsubsection*{TPIB-NN and VarTPIB-NN systems}
A shallow two-layer MFNN  is used for both TPIB-NN and VarTPIB-NN systems.
{Based on the best SER obtained on the development dataset, 34 hidden nodes were kept in the first layer while the second layer was set to 19 hidden nodes.}
The \textit{tanh} activation function is used in the first hidden layer, whereas the second layer is kept linear. 
We use the PyTorch\footnote{https://pytorch.org/} \cite{pytorch} implementation of stochastic gradient descent (SGD) with cross-entropy loss to train the MFNN models.
All the weights of MFNN are initialized using the initialization technique described in \cite{glorot2010understanding}.

\subsubsection*{TPIB-LDA and VarTPIB-LDA systems}
Only the first pass of  TPIB-LDA and  VarTPIB-LDA systems use a fixed number of clusters as a stopping criterion.
For both these systems, the first pass IB clustering is stopped when the number of clusters reaches 20. 
The second pass uses NMI as the stopping criterion.

\subsubsection*{Real Time Factor}
All RTFs are calculated on 2.6 GHz CPU with two threads without using GPUs.
The reported RTFs are calculated by averaging the RTFs across ten independent runs.

\begin{figure}[]
\centering
\includegraphics[width=0.37\textwidth]{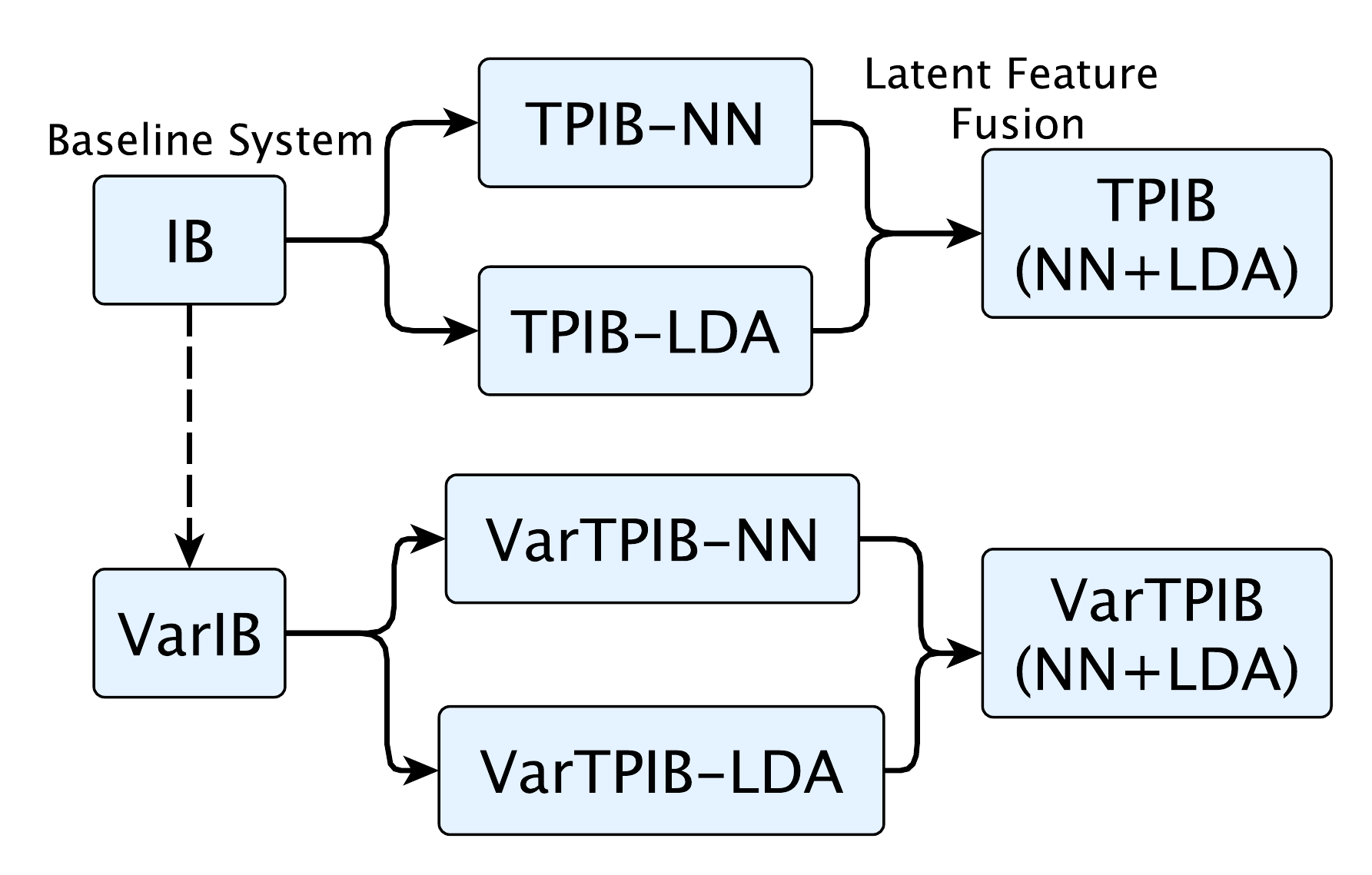}
\caption{Flow diagram for experiment protocol. IB system is used as the baseline.
TPIB/VarTPIB systems have NN and LDA variants. Fusion of latent features from NN and LDA is also studied.}
\label{fig:exp_flow}
\end{figure}

\subsubsection*{Experiment protocol}
{The complete flow of experiment design is shown in Fig. {\ref{fig:exp_flow}}.
The IB system is used as a baseline. 
The performance of TPIB-NN and TPIB-LDA systems are compared with IB system individually.
We have also performed an experiment with the fusion of latent features from NN and LDA models.
This denoted as TPIB (NN+LDA) in Fig. {\ref{fig:exp_flow}}.
The performance of the VarIB system is compared with the IB system.
Similar to the IB setup, experiments are performed on VarTPIB-NN, VarTPIB-LDA, and VarTPIB (NN+LDA)  systems.}

All the results reported in this paper are based on the best performing hyperparameters tuned on the development dataset.

\subsection{Results, Analysis, and Discussion}
\subsubsection{Speaker Error Rate} ~\\
The results of all the experiments in terms of SER are given in Table \ref{tab:ser}. 
The performance in terms of SER for all the proposed systems is compared with the baseline IB system.
All the systems use MFCC as an input in the first pass.
Speaker discriminative models and Latent Features (LF) used before the second pass are also mentioned in the table.

It can be seen in Table \ref{tab:ser} that the VarIB system performs better than the fixed length initialization in the baseline IB system for all datasets.
The VarIB system shows a 2.8\% absolute improvement on the development set.
Best-case absolute improvement of 1.5\% and 1.1\% is observed on RT-04Eval and RT-05Eval datasets, respectively.
Consistent improvements are observed across all datasets.
This suggests that equal distribution of speaker information in terms of phonemes across different segments leads to a better clustering solution. 
The other set of systems, including TPIB-NN and TPIB-LDA, consists of those that use the fixed length segment initialization.
Absolute improvements of 2.2\% and 3.2\% are observed on RT-05Eval dataset for TPIB-NN and TPIB-LDA systems, respectively.
It can be seen that both systems outperform the baseline IB system on most datasets.

\begin{figure}[t]
\centering
\includegraphics[scale=0.52]{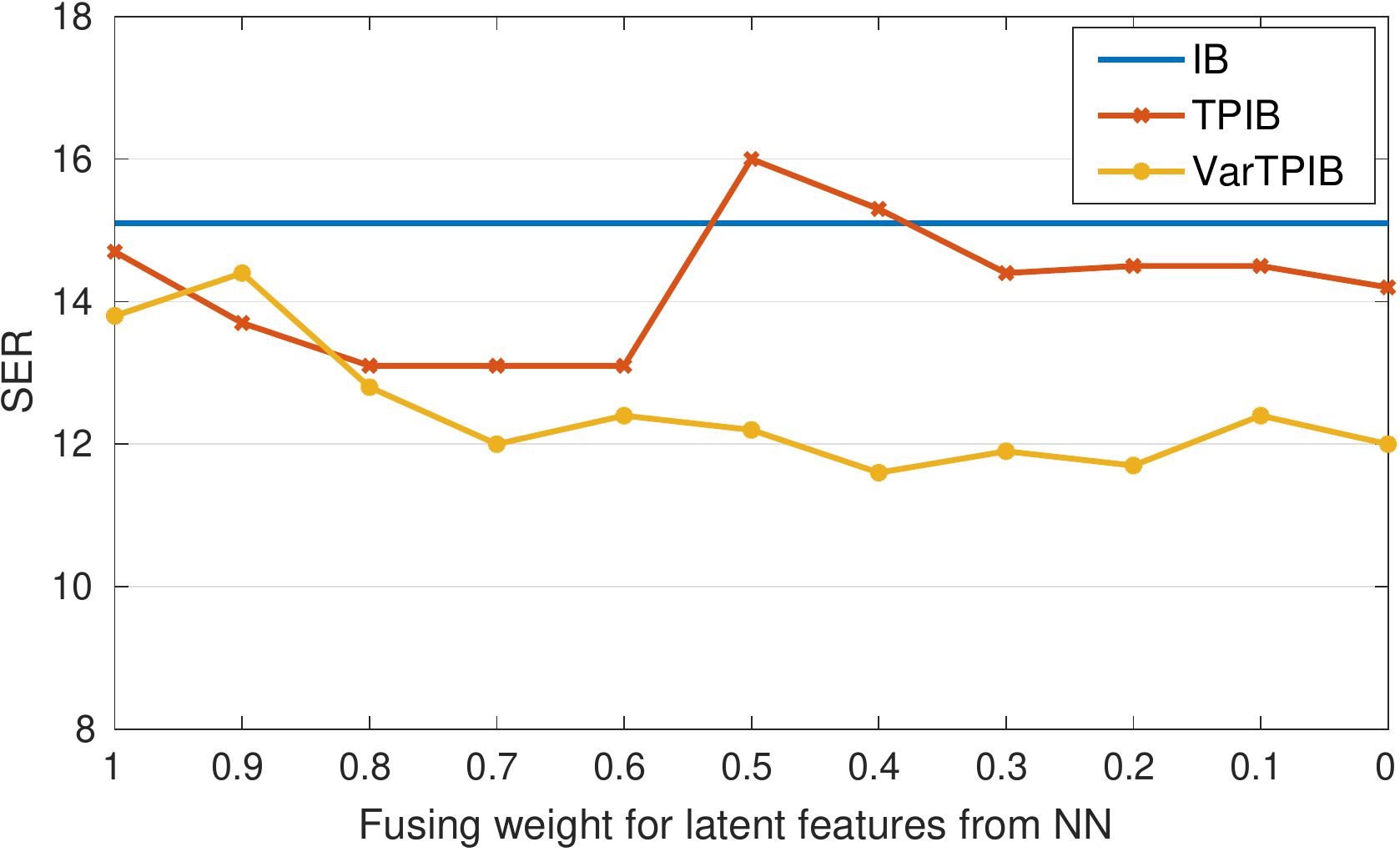}
\caption{SER for different feature fusing weights for $LF_{NN}$ and $LF_{LDA}$. X-axis denotes weight for $LF_{NN}$ (i.e. $w_N$) and the remaining $w_L=1-w_N$ is assigned to $LF_{LDA}$. 
The SERs are shown on the development dataset for different systems. The blue color line denotes SER of the IB based system.}
\label{fig:fuse}
\end{figure}

We also conduct an experiment where both MFNN and LDA are used after the first pass as speaker discriminative models.
Both the models were trained independently using the first pass output labels and the spectral features.
Since performing LDA is fast, the overall RTF of the TPIB-NN system does not change significantly.
The posteriors obtained from MFNN and LDA using $LF_{NN}$ and $LF_{LDA}$ were fused before the second pass as follows \cite{deepu11-fuseTDOA},

\begin{equation} 
	\label{eq:fuse}
P_t(\mathbf{y}) = P(\mathbf{y|F_t}).w_N + P(\mathbf{y|L_t}).w_L
\end{equation}
where $\mathbf{F_t}$ and $\mathbf{L_t}$ are feature vectors at time $\mathbf{t}$ from $LF_{NN}$ and $LF_{LDA}$, respectively.
{$\mathbf{y}$ is a vector representation for $y_i$ ($i=1\dots N$) and $y_i$ refers to the $i$-th Gaussian component of the respective feature streams.}
Here, $w_N$ and ${w_L}$ are the weights assigned to  $LF_{NN}$ and $LF_{LDA}$, respectively, such that ${w_N}$ + ${w_L}$ = 1.
The final posterior representation $P_t(\mathbf{y})$ is then used for the second pass IB clustering.
The combination of features shows improvement over individual latent feature streams on most datasets.
It shows absolute improvements of 3.8\% and 2.5\% compared to the baseline IB system on RT-05Eval and AMI-1 datasets, respectively.
This indicates the presence of complementary information in the latent features derived from different discriminative models.

In order to test the robustness of the proposed system against different feature weight combinations, we also report the average SER across all the feature fusing weights.
It can be seen that on average (under different feature weight combinations), the proposed systems perform significantly better than the baseline IB system on all datasets.
The best-case absolute improvement observed is 1.9\% on RT-05Eval dataset.
Notice that the SERs are averaged over different fusing weights and hence the reported improvement is significant.

The final set of proposed systems are VarTPIB-NN and VarTPIB-LDA. 
The VarTPIB-NN showed an improvement of 3.6\% and 2.6\% on RT-04Eval and AMI-2 datasets, respectively.
The VarTPIB-LDA showed an improvement of 3.9\% on RT-05Eval dataset.
The combination of latent features from MFNN and LDA results in a significant improvement of 4.7\% on AMI-1 dataset.
The latent feature combination in the case of VarTPIB systems did not always show an improvement over both the individual systems.
However, the observed SERs are always better than the baseline IB, TPIB, and VarIB systems on most datasets.
Similar to the systems using fixed length segment initialization (TPIB), the averaged SER over all the fusing weights for the VarTPIB system is better than the baseline IB on all datasets. 
It shows a significant absolute improvement of 2.5\% and 2.8\% over the baseline IB system on RT-05Eval and AMI-1 datasets, respectively.
As shown in Table \ref{tab:ser}, the SERs of VarTPIB systems are better than that of  VarIB and TPIB systems, 
showing clearly that the varying length segment initialization in tandem with a two-pass setup improves the performance significantly.

Fig. \ref{fig:fuse} shows SERs for different values of $w_N$ and $w_L$ weights for fixed length (TPIB's $LF_{NN}$ and $LF_{LDA}$) and varying length (VarTPIB's $LF_{NN}$ and $LF_{LDA}$) systems on development dataset.
It can be seen that both the systems perform better than the baseline for most of the values of fusing weights.
This confirms the robustness of all the proposed systems to feature fusing weight combinations.
It can be seen that the VarTPIB system performs consistently best among all the three systems.\\

\begin{table}[]
\centering
\caption{Comparison of proposed system with the state-of-the-art AHC on x-vectors with VB realignment. 
{SERs on best performing proposed system, i.e., VarTPIB (0.6, 0.4) is mentioned.}  
Lowest SERs achieved by the proposed approaches is also mentioned in the last column.}
\resizebox{\linewidth}{!}{
\begin{tabular}{ccccc}
\toprule
Dataset                                                       & \begin{tabular}[c]{@{}c@{}}Baseline IB\\ system\end{tabular} & \begin{tabular}[c]{@{}c@{}}State-of-the-art\\ x-vector system\end{tabular} & \begin{tabular}[c]{@{}c@{}}VarTPIB\\ (0.6, 0.4)\end{tabular} & \begin{tabular}[c]{@{}c@{}}Best SER on\\ proposed systems\end{tabular}    \\ \midrule

\begin{tabular}[c]{@{}c@{}}RT-04Dev\\ (Dev. set)\end{tabular} & 15.1                                                         & 10.4                                                              & 11.6                                                                          & 11.6                                                                  \\ \midrule
\multicolumn{5}{c}{Test Set}                                                                                                                                                                                                                                                    \\ \midrule
RT-04Eval                                                     & 13.5                                                         & 10.9                                                                    & 11.7                                                                    & 9.9                                                                 \\ 
RT-05Eval                                                     & 16.4                                                         & 10.4                                                               & 15.1        & 12.5                                                                  \\ 
AMI-1                                                         & 17.9                                                         & 9.7                                                                    & 13.2     & 13.2                                                                  \\ 
AMI-2                                                         & 23.5                                                         & 10.5                                                               & 21.1        & 20.9                                                                 \\ \bottomrule
\end{tabular}
}%
\label{tab:sota-ser}
\end{table}

\begin{table*}[t]
\centering
\caption{Module-wise RTFs (x10) for different systems on development dataset.}
\label{tab:rtf_all}
\resizebox{\textwidth}{!}{
\begin{tabular}{ccccccccc}
\toprule
System   & \begin{tabular}[c]{@{}c@{}}Phoneme\\ Boundaries \end{tabular} & \begin{tabular}[c]{@{}c@{}}Posterior\\ Computation 1\end{tabular} & \begin{tabular}[c]{@{}c@{}}IB \\ Clustering 1\end{tabular} & \begin{tabular}[c]{@{}c@{}}KL-HMM \\ Realignment 1\end{tabular} & \begin{tabular}[c]{@{}c@{}}Discriminative \\ Model Training\end{tabular} & \begin{tabular}[c]{@{}c@{}}Posterior \\ Computation 2\end{tabular} & \begin{tabular}[c]{@{}c@{}}IB \\ Clustering 2\end{tabular} & \begin{tabular}[c]{@{}c@{}}KL-HMM \\ Realignment 2\end{tabular} \\
\midrule
IB     & -  & 0.19                                                              & 0.46                                                       & 0.09                                                            &  -                                                                       & -                                                                  & -                                                          & -                                                               \\
VarIB     & 0.2  & 0.17                                                           & 0.36                                                       & 0.09                                                            &  -                                                                       & -                                                                  & -                                                          & -                                                               \\
TPIB-NN  & - & 0.17                                                              & 0.50                                                       & 0.09                                                            & 0.97                                                                     & 0.16                                                               & 0.46                                                       & 0.09                                                            \\
TPIB-LDA & - & 0.18                                                              & 0.48                                                       & \textbf{0}                                                               & \textbf{0.01                                                                    } & 0.18                                                               & 0.50                                                       & 0.07         \\
\bottomrule                                                  
\end{tabular}
}
\end{table*}

\textit{Comparison of proposed systems with the state-of-the-art x-vector based system:} 

In this section, we compare the performance of the proposed systems with the state-of-the-art  Agglomerative Hierarchical Clustering  (AHC) \cite{ahc} on x-vector (deep speaker embedding) system \cite{sell-dihard2018}.
This system clusters the x-vector embeddings in an agglomerative bottom-up approach.
Note that the x-vector model is trained on huge amounts of labeled data in a supervised manner, whereas the proposed systems (and the baseline IB system) do not use any speaker labeled data.

We follow the open-source Kaldi recipe for building the x-vector system \cite{sell-dihard2018}. 
It uses around 5k hours of data (including clean speech and augmentation data) for training the time delay neural network (TDNN) \cite{tdnn} in a supervised fashion.
The hyperparameters for x-vector extraction, i.e., the window length and the period, are kept at 3.5 seconds and 1.75 seconds, respectively.
Probabilistic Linear Discriminant Analysis (PLDA) {{\cite{plda-prince}~\cite{plda-sell}}} models are whitened using corresponding NIST and AMI datasets.
The PLDA score threshold is also calibrated on the development dataset.
The speaker boundaries obtained after AHC on x-vector is subjected to Variational Bayes (VB) realignment \cite{vb-zheng}. 
Similar to \cite{sell-dihard2018}, the parameters for VB are learned using the VoxCeleb dataset.  
All the hyper-parameters are tuned to give the best results on the development dataset.

As expected, the x-vectors based diarization system performs significantly better than the unsupervised baseline IB system on all datasets.
It can be seen from the Table {\ref{tab:sota-ser}} that the best system, i.e., the VarTPIB system with fusing weights of (0.6, 0,4) show comparable performance to the x-vector system on some datasets (RT-04Dev and RT04Eval).
The lowest SER performance (last column) of the proposed systems shows how close the proposed systems can reach near to the x-vector system's performance.
It can be seen that the best SERs are comparable to that of the state-of-the-art x-vector system on most datasets (except for AMI-2 dataset).

Detailed analysis on the performance related to the IB approach is given in \cite{deepu09-ib}.
The GMM in an IB approach is estimated using data from the same meeting. 
The variations in the background noise in a recording and the overlapping speech affect the IB clustering process \cite{deepu09-ib}.
Since the proposed systems are based on the IB approach, they show similar behavior.
The AMI-2 dataset contains noisy recordings that show high SER on the IB approach. 
For example, 
IS1006c and ES2013b meetings show a high SER of 40\% and 38\% for the IB system, respectively.
This is mainly attributed to the lack of robustness to the noisy conditions for IB approach.
On the other hand, the TDNN model in the x-vector system is trained on 1000s of hours of speaker labeled data that is augmented with noise datasets (MUSAN and RIRs)\footnote{https://openslr.org} hence it is robust to different types of noise.
Note that the proposed systems do not use any speaker labeled data to obtain speaker embeddings, unlike the x-vector based system.  \\

\begin{figure}[t]
\centering
\includegraphics[scale=0.53]{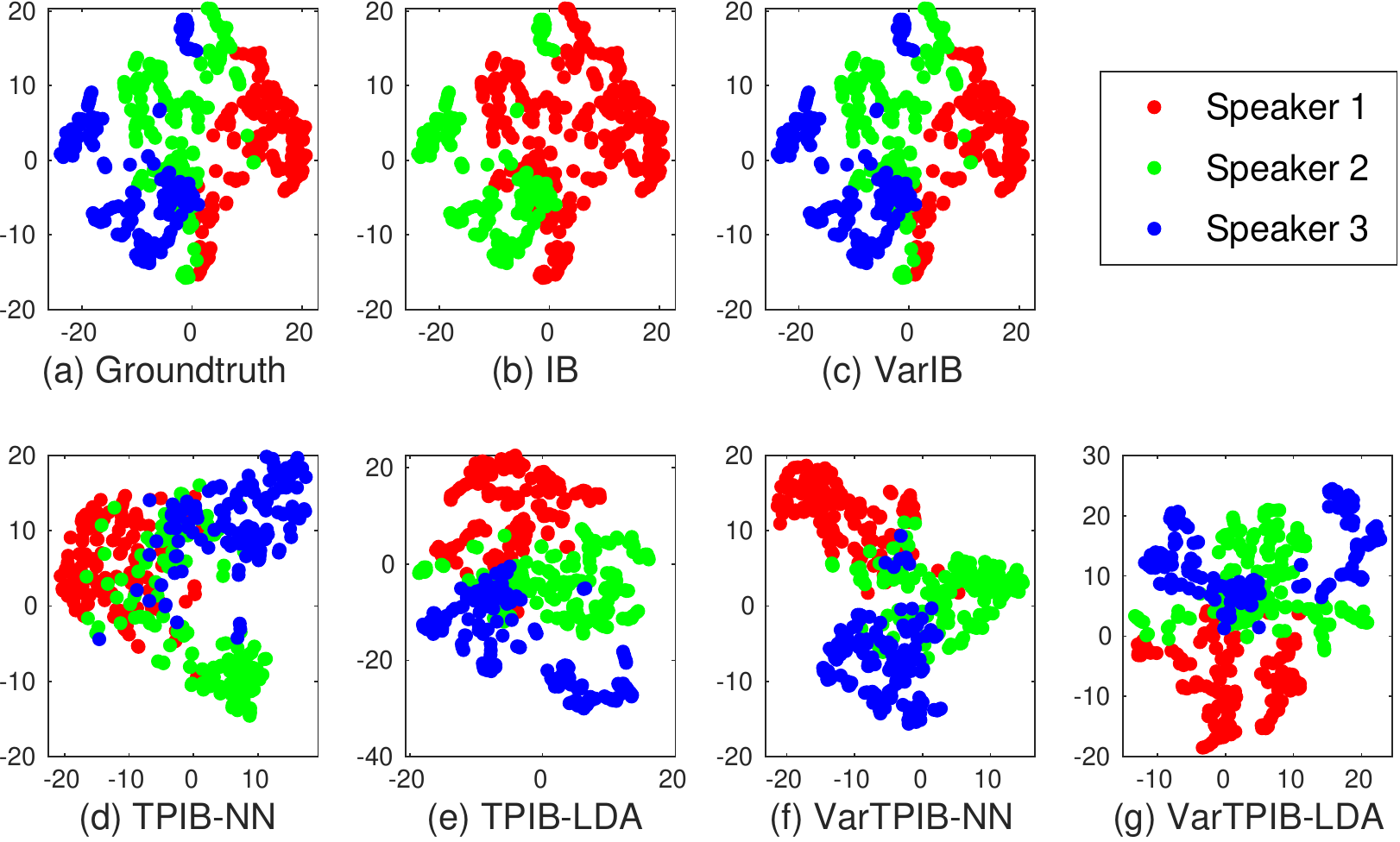}
\caption{t-SNE plots for segments in a recording from development set. Different colors represent different speakers.}
\label{fig:tsne}
\end{figure}

\subsubsection{On Speaker Discrimination} ~\\
To observe the speaker separation and the correctness in the speaker's relative labels, we plot the t-Distributed Stochastic Neighbor Embedding (t-SNE) projections \cite{tsne} of 3 speaker segments along with their speaker identities (relative) taken from all systems.
For the t-SNE study, the actual speaker identities corresponding to the ground-truth speaker labels were obtained using the standard NIST evaluation tool. 
This was cross-verified by listening to the segments for the correctness.
Fig. \ref{fig:tsne} shows the t-SNE projections on the 2d plane for short speaker segments in different feature spaces for all systems. 
Fig. \ref{fig:tsne}(a) shows the ground truth in MFCC space where three different speakers are present.
Fig. \ref{fig:tsne}(b) also shows the MFCC space, but IB system has detected only two speakers.
The VarIB correctly identifies three speakers in the MFCC space itself, as it can be seen from Fig. \ref{fig:tsne}(c).
The t-SNE plot for  TPIB-NN system (Fig. \ref{fig:tsne}(d)) shows that the three speakers are correctly detected, but the speaker spaces do overlap.
The VarTPIB-NN system shows a better separation than TPIB-NN as seen in Fig. \ref{fig:tsne}(f).
Similar trends were also observed for $LF_{LDA}$ space in the case of TPIB-LDA (Fig. \ref{fig:tsne}(e)) and VarTPIB-LDA (Fig. \ref{fig:tsne}(g)).
Overall, the proposed discriminative frameworks emphasize on the differences between speakers in a given conversation.

\subsubsection{Real Time Factor} 
\label{subsubsec:rtf} ~\\
The module-wise split in RTFs scaled by a factor of 10 (for readability) for IB, VarIB, TPIB-NN and TPIB-LDA systems are shown in Table \ref{tab:rtf_all}.
The VarTPIB-NN and VarTPIB-LDA systems show similar trends in module-wise RTFs as TPIB-NN and TPIB-LDA, respectively.
Hence, we do not report the module-wise split in RTF for VarTPIB-NN and VarTPIB-LDA systems.
It can be seen that the significant part of the RTF, i.e., 0.97 in TPIB-NN system comes from MFNN training, which is reduced significantly to 0.01 for TPIB-LDA system.
Early stopping of the first pass and no KL-HMM realignment after the first pass also contributes to the reduction in runtime for TPIB-LDA.

\begin{table}[t]
\centering
\caption{Overall RTF (x10) for different systems on development dataset.}
\label{tab:rtf}
\resizebox{0.25\textwidth}{!}{
\begin{tabular}{ccc}
\toprule
 System     	 & RTF (x10) \\ \midrule 
IB   		  & 0.74  \\  \midrule
x-vector  & 2.13 \\  \midrule
VarIB 		 & 0.82 \\ \midrule
TPIB-NN    &  2.44 \\
TPIB-LDA    &  1.42 \\ \midrule
VarTPIB-NN   &  2.58 \\
VarTPIB-LDA   &  1.61 \\
\bottomrule
\end{tabular}}
\end{table}

The overall RTFs scaled by a factor of 10 for different systems are given in Table \ref{tab:rtf}.
The RTF of x-vector based system during test time is 2.13, which includes the runtime of x-vector extraction from the trained TDNN model, PLDA score estimation, performing AHC on x-vectors, and VB based refinement. 
The IB and VarIB systems are the fastest among all the systems with a similar RTF of 0.74 and 0.82, respectively.  
The VarIB system improves the SER without significantly degrading the RTF of IB system.
The RTF of TPIB-NN and VarTPIB-NN systems are in a similar range of 2.4 - 2.6. 
Note that the given RTFs are scaled by a factor of 10.
The RTF of TPIB-LDA and VarTPIB-LDA systems are also in a similar range of 1.4 - 1.6. 
The TPIB-LDA and VarTPIB-LDA systems show 41.8\% and 37.6\% relative improvements in RTF compared to TPIB-NN and VarTPIB-NN systems, respectively.

Overall, the proposed systems have RTFs well below 1 (or 10 in scaled RTF value) and are only 2-3 times slower than  IB based system with significant improvements in SER.

\section{Summary and Conclusion}
\label{sec:con}
We propose different architectures for IB based diarization system focusing on two critical challenges in diarization; (i) Segment initialization, and (ii) Speaker discriminative representation.
In the first part, we propose a varying length segment initialization technique for the IB system using phoneme rate information.
Here, the segments are initialized such that different segments have a similar number of phonemes. 
We show that a proper initialization of segments results in a better SER with negligible degradation in runtime.
In the second part of the work, we show how the TPIB-NN and TPIB-LDA systems help to incorporate the speaker discriminative features during the diarization process itself.
This in turn improves the diarization performance significantly. 
Moreover, the runtime of TPIB-NN is further improved by the TPIB-LDA system.
In the third part of the work, we combine the VarIB and TPIB systems to propose VarTPIB systems that leverage the advantages of both, i.e., better segment initialization and speaker discriminative features.
The VarTPIB systems show significant improvement in the performance compared to the baseline IB system, and the proposed VarIB and TPIB systems. 
In conclusion, we show that a good segment initialization along with the speaker discriminative features results in a good clustering solution.

As the focus of this work is towards different frameworks, we used standard MFCC in all our experiments.
However, other spectral features like MFS and LFS {\cite{dawalatabad16-tisd}} {\cite{srikanth14-mfs}}, and  group delay based features {\cite{padmanaban-gd}}  can also be used in the proposed architectures. 
The proposed systems are generic enough to be used along with deep embeddings instead of standard spectral features.
Hence, in future we would also like to study different data-derived features such as i-vectors {\cite{srikanth15-ivec}},  d-vectors {\cite{dvector-1}}, x-vectors {\cite{xvector}}, and 
 c-vectors {\cite{cvector}}  in the proposed architectures.
We also plan to extend the TPIB and the VarTPIB systems to perform multiple passes of IB and VarIB based clustering to further refine the speaker boundaries.

\section*{Acknowledgement}
\label{sec:ack}
We would like to thank the Defence Research and Development Organisation (DRDO), India for supporting a part of this work under the project CSE1314142DRDOHEMA.

%

\bibliographystyle{IEEEtran}
\bibliography{ref}

\end{document}